\documentclass[prb,aps,superscriptaddress,twocolumn]{revtex4}
\usepackage{graphicx}% Include figure files
\usepackage{dcolumn}% Align table columns on decimal point
\usepackage{bm}% bold math

\usepackage{amsmath, amssymb, graphics, setspace}

\usepackage{color} %\YS required for the textcolor command

\newcommand{\mathsym}[1]{{}}
\newcommand{\unicode}[1]{{}}

\begin{document}

\title{Importance of anisotropic exchange interactions in honeycomb iridates. Minimal model for zigzag antiferromagnetic order in Na$_2$IrO$_3$.}

\author{Yuriy Sizyuk}
\affiliation{Department of Physics, University of Wisconsin,
Madison, Wisconsin 53706, USA}
\affiliation{School of Physics and Astronomy, University of Minnesota, Minneapolis,
MN 55116, USA}

\author{Craig Price}

\affiliation{Department of Physics, The Pennsylvania State University, 104 Davey Lab, University Park, Pennsylvania 16802, USA}

\author{Peter W\"{o}lfle}
\affiliation{Department of Physics, University of Wisconsin,
Madison, Wisconsin 53706, USA}

\affiliation{Institute for Condensed Matter Theory and Institute for Nanotechnology, Karlsruhe Institute of Technology, D-76128 Karlsruhe, Germany}

\author{Natalia  B. Perkins}
\affiliation{Department of Physics, University of Wisconsin,
Madison, Wisconsin 53706, USA}

\affiliation{School of Physics and Astronomy, University of Minnesota, Minneapolis,
MN 55116, USA}

\begin{abstract}
In this work, we investigate the microscopic nature of the magnetism in  honeycomb iridium-based systems
by performing a systematic study of  how the effective magnetic interactions in these  compounds
 depend on various electronic microscopic parameters.
We show that the minimal model describing  the magnetism in A$_2$IrO$_3$ includes both isotropic and anisotropic Kitaev-type spin-exchange interactions between nearest and next-nearest neighbor  Ir ions, and that the magnitude of the Kitaev interaction between next-nearest neighbor Ir  magnetic moments is comparable with nearest neighbor interactions. We also find that, while the Heisenberg and the Kitaev interactions between nearest neighbors are correspondingly  antiferro- and ferromagnetic,  they both change sign for the next-nearest neighbors.
Using classical Monte Carlo simulations we examine the magnetic phase diagram of the derived  super-exchange model.
Zigzag-type antiferromagnetic order is found to occupy a large part of the phase diagram of the model and, for ferromagnetic next-nearest neighbor Heisenberg interaction relevant for Na$_2$IrO$_3$, it can be stabilized  even  in the absence of third nearest neighbor coupling.
Our results suggest that a natural physical origin of the zigzag phase experimentally observed in Na$_2$IrO$_3$ is due
to the interplay of the Kitaev  anisotropic interactions between nearest  and next-nearest neighbors.
 \end{abstract}

\maketitle

\section{Introduction}

The  magnetism in $4d$ and $5d$ transition metal (TM) oxides, particularly  realized in  iridates and rhodates,
 has  recently attracted a lot of interest. In these systems, the interplay between the spin-orbit (SO) coupling, the crystal field (CF) splitting,  and Coulomb and Hund's coupling  leads to
 a rich variety of magnetic exchange interactions, new types of magnetic ground states and excitations.

  In our recent  work\cite{perkins14} (hereafter referred to as paper I), we  applied the Mott insulator scenario,
   extending the original study by Jackeli and Khalliulin,~\cite{jackeli09} and  developed  a theoretical framework   for the derivation of effective super-exchange Hamiltonians  that govern the  magnetic properties of  systems with strong SO coupling.  In our approach,  both the  many-body (Coulomb and Hund's interaction) and the single electron (SO and CF interactions) effects are treated on an equal footing. In this framework, we first determined
 the  localized degrees of freedom of the iridium system by finding the exact eigenstates of the single-ion  microscopic  Hamiltonian   for Ir$^{4+}$  ions, and  then computed the interactions between them.   Because of  time reversal symmetry of the single-ion Hamiltonian, the lowest  atomic state is always at least two-fold degenerate, and can be described using pseudospin-$1/2$ operators.

In paper I, some of us showed that   the super-exchange Hamiltonian describing  interactions between these pseudospins might have unusual anisotropic components. Moreover, these anisotropic interactions might be the dominating   interactions between magnetic moments. The form of these anisotropic interactions may also be quite unusual.  In particular, they do  not need to be confined to the traditional anisotropic interaction types acting equally on all sites of the lattice (i.e. easy-plane or easy-axis anisotropy).
 Instead, the anisotropic interactions might involve coupling between different components of spins sitting on different lattice sites.
 The Dzyaloshinskii-Moriya  interaction\cite{dzyalo58,moriya60}  and  the Kitaev interaction on the honeycomb lattice\cite{kitaev06,jackeli10} are salient examples of such interactions.

\begin{figure}
\label{fig1}
\includegraphics[width=0.95\columnwidth]{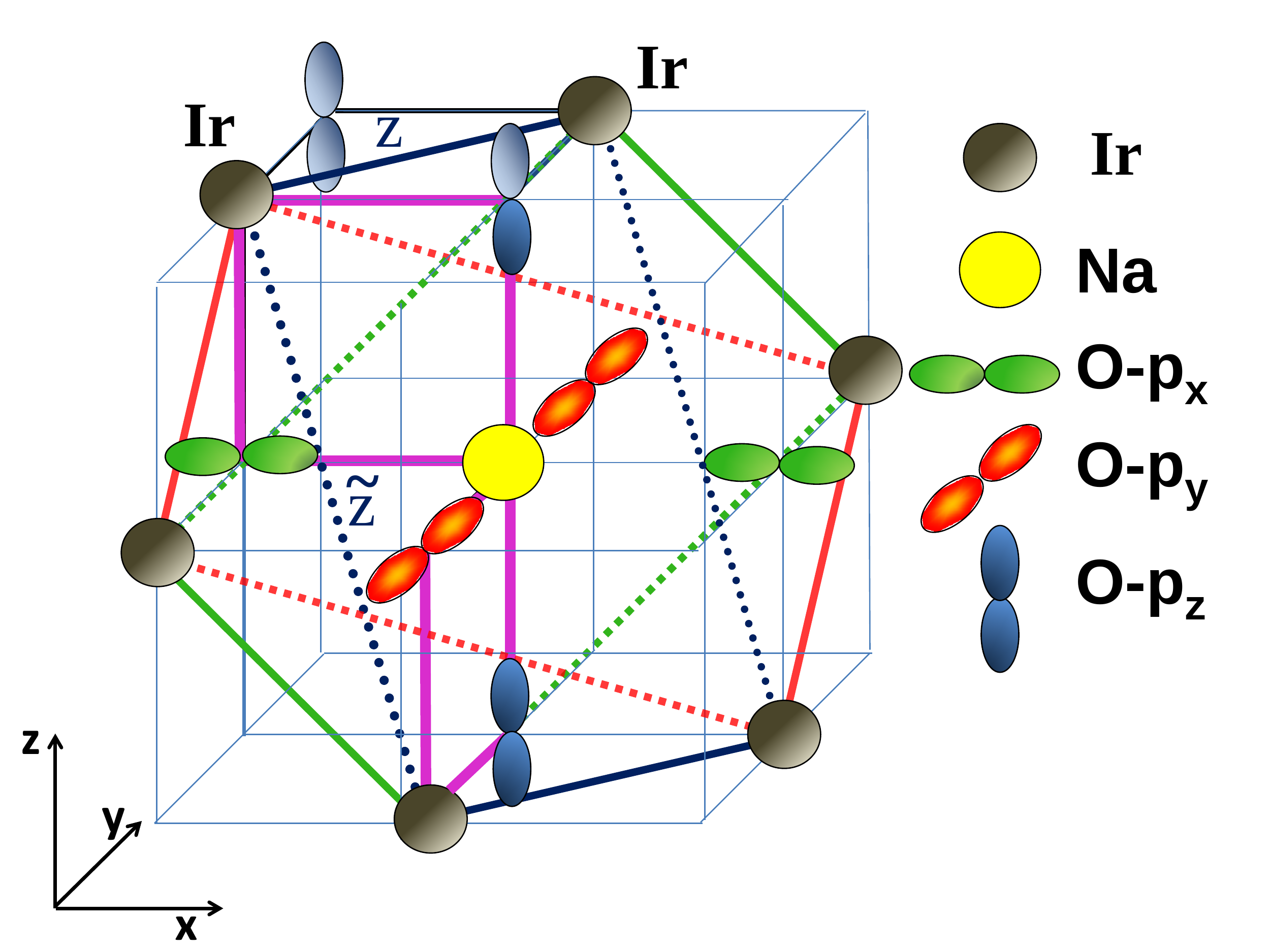}
\includegraphics[width=0.85\columnwidth]{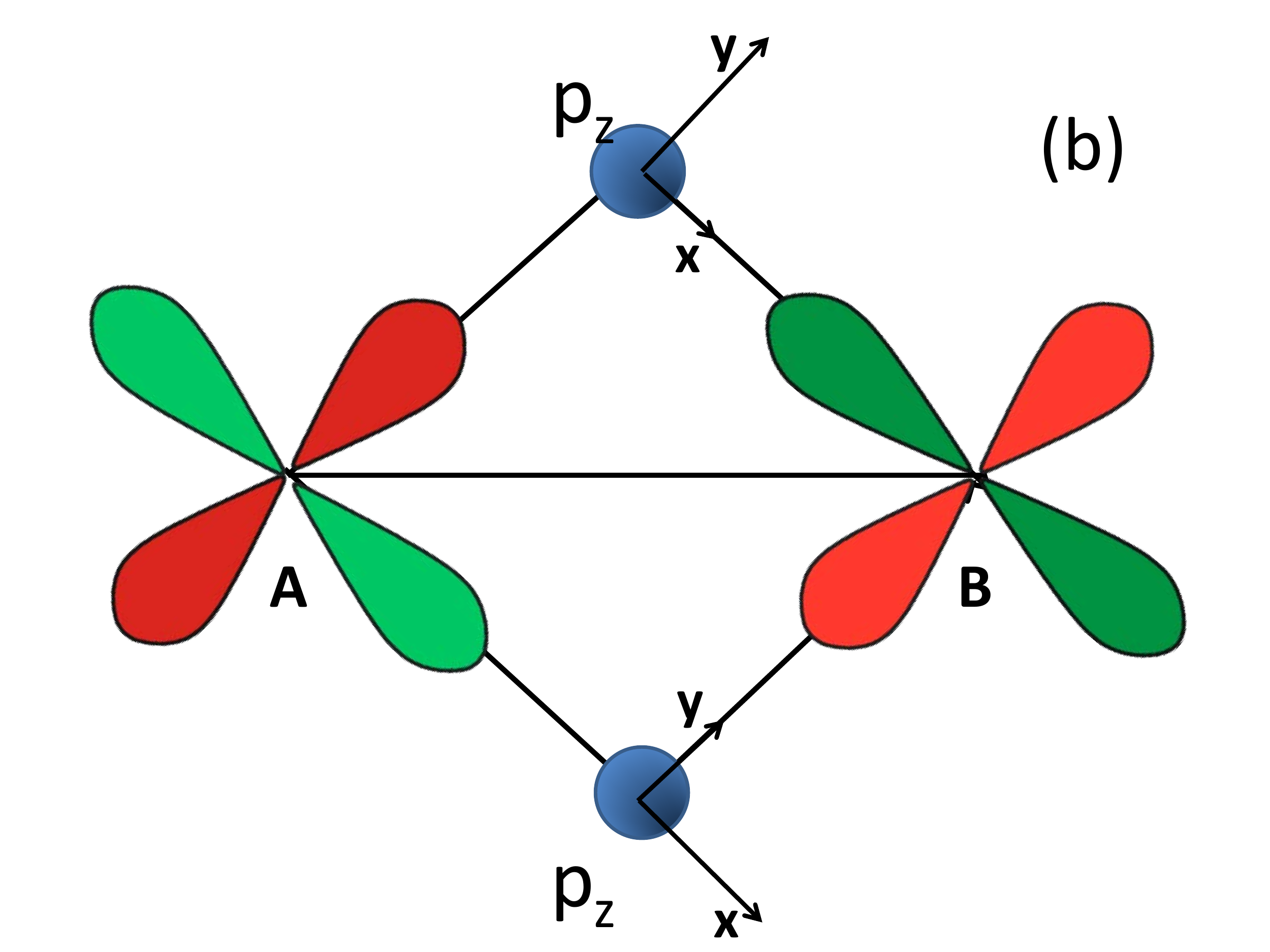}
\caption{(Colors online) (a) Schematic representation of A$_2$IrO$_3$ structure.
 $x-$, $y-$ and $z-$ n.n. Ir-Ir bonds  are shown by   red, green and blue solid lines. ${\tilde x}-$, ${\tilde y}-$ and ${\tilde z}-$ second n. n. Ir-Ir bonds  are shown by   red, green and blue dotted lines. Thick magenta lines represents Ir-O-Na-O-Ir second n. n. super-exchange paths.
(b)
Undistorted 90$^\circ$ Ir-O-Ir bond. Local axes for Ir$^{4+}$ ions on A  and  B
sublattices are the same as the global axes. Two possible super-exchange paths via upper or lower oxygen are shown.}
\end{figure}

 In paper I, we focused on  iridates and rhodates  with tetragonal symmetry, e.g. we studied in detail the magnetic interactions in Sr$_2$IrO$_4$.~\cite{cao98,bjkim08,moon09,kim09,comin12,fujiyama14} Our approach allowed us to  show that the weak coplanar ferromagnetism observed in Sr$_2$IrO$_4$\cite{cao98} is governed by the Dzyaloshinskii-Moriya interaction with an unusual strength owing to the  large SO coupling.
In the present paper,  we make use  of the  experience obtained in paper I to study the magnetic properties of A$_2$IrO$_3$~\cite{singh10,singh12,liu11,ye12,choi12}  (A=Na,Li) in which  the Ir$^{4+}$  ions occupy the sites of a honeycomb  lattice.

The nearest-neighbor (n. n.) super-exchange in honeycomb iridates in the absence of lattice distortions, the so-called Kitaev-Heisenberg (KH)  model, was first proposed by Jackeli and Khalliulin.~\cite{jackeli09,jackeli10}
They showed that in these  systems the coupling between n. n. Ir magnetic moments occurs  through  both direct exchange between  Ir$^{4+}$ ions and  through  a super-exchange coupling mediated by an intermediate oxygen along the   90$^{\circ}$ Ir-O-Ir bond.   The latter process gives rise to a nonzero  anisotropic interaction between pseudospins, which has the form of the aforementioned Kitaev interactions,  but only for a finite value of the Hund's coupling.
The KH model correctly captures the nature of the anisotropic part of the  magnetic interactions in   Na$_2$IrO$_3$  honeycomb compounds and also predicts some non-trivial properties of these compounds at finite temperatures.~\cite{price12,price13}  Nevertheless, the model does miss some essential features: it does not account for  both the  zigzag magnetic order and for the spectrum of magnetic excitations in Na$_2$IrO$_3$ measured in neutron scattering experiments.~\cite{liu11,ye12,choi12}   Partly, this is because the original KH model
 neither includes further neighbor interactions, which have been shown to play a significant role in stabilizing  the zigzag antiferromagnetic ordering in Na$_2$IrO$_3$,~\cite{kimchi11,choi12}
  nor lattice distortions, which might  also be essential for these compounds.

 In this work we revisit  the KH model\cite{jackeli09,jackeli10} and derive its extension up  to second neighbor's interactions, starting from the exact eigenstates of the single-ion  microscopic Hamiltonian which equally includes both the SO coupling and the trigonal distortion. In this context, our work differs from the recent study by Bhattacharjee,  Lee and  Kim,~\cite{subhro12}
in which the effective spin Hamiltonian  was derived  by setting the energy scale associated with
trigonal distortion to infinity first, followed by that of the SO energy scale.
  Here, we estimate the strength of  magnetic interactions  in Na$_2$IrO$_3$ based on  the tight-binding parameters obtained from the ab-initio density-functional theory study by Foyevtsova {\it et al}.\cite{katerina13}
We show that the effective spin Hamiltonian on the honeycomb lattice, whose bonding geometry is shown in Fig. 1 (a),
 contains   several anisotropic spin interactions among which the strongest is the  Kitaev interaction between nearest neighbors.

 We also compute the super-exchange interaction between the second neighbors  forming two triangular sublattices, and find that it  is of  a form similar to the n. n. interaction, i.e. the dominant part can be written as a sum of  isotropic Heisenberg and  anisotropic Kitaev terms. These  interactions are only slightly smaller than the n. n. Kitaev interactions. Other anisotropic interactions,  which couple different components of spins on a given bond, are significantly smaller and most of them are non-zero only in the presence of trigonal lattice distortions. In this respect they are different from  the Kitaev-like interactions which are present  even in the ideal structure.

The magnetic phase diagram which emerges from our study is
presented in Fig. 5. This is the key result of this paper. We argue that the zigzag magnetic order, experimentally observed in Na$_2$IrO$_3$,   is stabilized by the interplay of four major interactions: isotropic {\it antiferromagnetic} and anisotropic {\it ferromagnetic} Kitaev interactions  for  n. n. bonds, isotropic {\it ferromagnetic} and  anisotropic {\it antiferromagnetic}  Kitaev interactions for the next-nearest  neighbors. Unlike in other theoretical studies of magnetic properties of Na$_2$IrO$_3$\cite{kimchi11,jackeli13,katukuri14}, in our model the zigzag phase is stabilized  for both  the correct signs of n. n. interactions, and  even without invoking third neighbor interactions.

The rest of the paper is organized as follows.
In Sec.~\ref{sec:model}, we  introduce
the single ion microscopic model appropriate for the description of the physical
properties of   iridates on the honeycomb lattice.
We  first obtain one-particle eigenstates  taking into account only SO coupling and trigonal CF interaction. We then  compute two-particle excited eigenstates   fully considering  correlation effects. Then, in Sec.~\ref{sec:ham}, we briefly review the derivation of an effective super-exchange Hamiltonian for these systems. All technical details of the derivation  can be found in our previous work.\cite{perkins14}
 In Sec.~\ref{sec:hop}, we obtain hopping matrices for neighboring
iridium ions.  Our calculation is  based on a tight-binding fitting
of ab-initio electronic structure in the presence of trigonal
distortion performed by Foyevtsova {\it et al}.\cite{katerina13}
In Sec.~\ref{sec:coupling}, we present our results on the magnetic interactions.  We show that these interactions can be most generally represented by
a $3\times 3$  bond-dependent exchange coupling matrix.  We show  that,  while the Kitaev-type of anisotropy is determined by the inequality of its diagonal elements due to the Hund's coupling, the off-diagonal matrix elements are anisotropies  mostly caused by the trigonal crystal field. In Sec.~\ref{sec:pd},
taking into account  only the dominant interactions,  we  perform  classical  Monte Carlo simulations and obtain the low-temperature phase diagram of the minimal super-exchange model for honeycomb iridates.
 We conclude in Sec.~\ref{sec:conclusion}
with a summary and discussion of our results.

\section{Single-ion Hamiltonian}\label{sec:model}

\subsection{One-particle eigenstates}

In all  iridates considered here,  the Ir$^{4+}$ ions sit inside an oxygen  cage forming an octahedron.  This octahedral CF splits the five 5$d$ orbitals of Ir$^{4+}$ into doubly degenerate $e_g$
orbitals at higher energy  and into the three-fold degenerate $t_{2g}$ multiplet.
 In iridates, the energy difference between $e_g$ and  $t_{2g}$ levels is large and is typically of the order 2-3 eV. Because of this, the five electrons occupy  only the low lying $t_{2g}$ orbitals. As a consequence, the on-site interactions, such as the SO, Coulomb  and Hund's interactions, as well as additional symmetry-lowering  CF interactions, e.g. the trigonal CF,  can be considered within the $t_{2g}$ manifold only.
In this limit of large octahedral CF, the  SO coupling has to be projected onto the $t_{2g}$  manifold, assuming an
effective orbital angular momentum $L=1$.
In terms of local axes, which are bound to the oxygen octahedron, the $t_{2g}$ orbitals  of Ir ions are
$\left\vert X\right\rangle \equiv\left\vert yz\right\rangle $,
$\left\vert Y\right\rangle \equiv\left\vert zx\right\rangle $, and
$\left\vert Z\right\rangle \equiv\left\vert xy\right\rangle $.
The SO  and trigonal CF  interactions give rise to a splitting of the levels according to the symmetry of the underlying lattice.
In the case of the  honeycomb  iridates, A$_2$IrO$_3$, the trigonal CF arises from a compression of the oxygen cages along the $[111]$ directions (local $C_3$ axis).  At ambient pressure, the splitting of the $t_{2g}$ levels due to the trigonal CF  is about 110 meV~\cite{gret13} which is smaller,  but of the same order of magnitude as the SO coupling, which is   about 400 meV. Therefore,  here we treat the SO coupling and the trigonal CF interactions on the same footing. Also,  it is believed that much larger values  of the trigonal distortion can be reached by applying uniaxial pressure.

   Since the Hamiltonian is time-reversal invariant, the ground-state of the single-ion single-hole ($5d^5$ configuration of Ir$^{4+}$ ion) is a Kramer's doublet, which we represent as a pseudospin-1/2.
  However, the choice of the two orthonormal states within the doublet that would represent the pseudospin-up and pseudospin-down states deserves some well-inspired consideration, as this choice determines the coordinate system of the final super-exchange Hamiltonian. Since the most prominent anisotropy, the Kitaev interaction, has the simplest form in the coordinate system bound to the cubic axes of the oxygen octahedron environment, we choose the two orthogonal states that correspond to this particular  Cartesian reference frame. In the absence of the trigonal distortion, the ground state doublet is simply a $J_{\rm eff}=1/2$ doublet and the good choice of the states within it are the $J_{\rm eff}^z=\pm1/2$ states. In the presence of the trigonal distortion, the choice of the representation  is not as straightforward since the ground state doublet contains a mixture of both $J_{\rm eff}=1/2$ and $J_{\rm eff}=3/2$ states.
  To resolve this, we first find a random set of orthonormal states within the doublet and then make linear combinations of them in such a way that pseudospin-1/2 "up-state" has no $\left\vert J_{\rm eff}=1/2,J_{\rm eff}^z=-1/2\right\rangle$ component, whereas pseudospin-1/2 "down-state" has no $\left\vert J_{\rm eff}=1/2,J_{\rm eff}^z=1/2\right\rangle$ component. Namely, we allow the trigonal CF to admix the $J_{\rm eff}=3/2$ states to the $J_{\rm eff}=1/2$ states, but we do not allow the latter to mix among themselves.

  In the most simple form, the single-ion Hamiltonian can be written  when the  axis of the quantization of angular momentum is along the $[111]$ direction:
\begin{eqnarray}\label{ham_CF-SOC}
H_{\lambda,\Delta}=\lambda {\mathbf S}\cdot{\mathbf L}
+\Delta L_{[111]}^{2},
\end{eqnarray}
where $L_{[111]}$ denotes the component of the angular momentum along the $[111]$ axis.
Here the first term describes the SO coupling and the second term describes the trigonal CF. However, this form is not useful  if we want to obtain our final result in the  Cartesian reference frame bounded to the cubic crystallographic axes.   If now we rewrite the CF term in terms of it's eigenstates, then the Hamiltonian (\ref{ham_CF-SOC}) becomes:
\begin{eqnarray}\label{ham_CF-SOC2}
H_{\lambda,\Delta}=\lambda {\mathbf S}\cdot{\mathbf L}
+\frac{\Delta}{3} \left(-2 |a_{1g}\rangle\langle a_{1g}|+|e_g^+\rangle \langle e_g^+|+|e_g^-\rangle \langle e_g^-|\right),
\end{eqnarray}
where  the crystal field eigenstates include the low-energy singlet $|a_{1g}\rangle$ and the higher energy doublet
$|e_g^\pm\rangle$. The singlet state can be written as
\begin{eqnarray}\label{eq:a1g}
    |a_{1g}\rangle = \hat\nu_x |X\rangle
    + \hat\nu_y |Y\rangle + \hat\nu_z |Z\rangle,
\end{eqnarray}
where $\hat{\bm\nu}=(\hat\nu_x, \hat\nu_y,\hat\nu_z)$  is the unit
vector parallel to the $[111]$ trigonal axis ($ \hat\nu_j=1/\sqrt{3}$). The doublet state  can be conveniently written using the following chiral basis:
\begin{eqnarray}
    \label{eq:eg}
    \begin{array}{c}
    |e_g^+\rangle  = \hat\nu_x e^{-i\omega} |X\rangle +
    \hat\nu_y e^{+i\omega} |Y \rangle + \hat\nu_z |Z\rangle, \\
    |e_g^-\rangle  = \hat\nu_x e^{+i\omega} |X\rangle +
    \hat\nu_y e^{-i\omega} |Y \rangle + \hat\nu_z |Z\rangle,
    \end{array}
\end{eqnarray}
where $\omega \equiv 2\pi/3$. Now, that the CF part of the Hamiltonian is written in an $L$-independent way, we are free to choose the angular momentum quantization axis along the cubic $z$ direction for our basis. The basis we use is $
{\hat J}=\{
|\frac{1}{2},\frac{1}{2}\rangle,
|\frac{1}{2},-\frac{1}{2}\rangle,
|\frac{3}{2},\frac{3}{2}\rangle,
|\frac{3}{2},\frac{1}{2}\rangle ,
|\frac{3}{2},-\frac{1}{2}\rangle,
|\frac{3}{2},-\frac{3}{2}\rangle\}
$. The details of this basis and its relation to the basis of the cubic orbitals are given in paper I.\cite{perkins14} The Hamiltonian matrix in this basis is given by
\begin{widetext}
\begin{eqnarray}\label{HAM}
 {\hat H} =
 \left( \begin{array}{cccccc}
-\lambda & 0 & -\frac{(1-\imath)\Delta}{3\sqrt{6}} & 0 & \frac{(1+\imath)\Delta}{3\sqrt{2}} & \frac{\imath\Delta}{3}\sqrt{\frac{2}{3}} \\
0 & -\lambda & \frac{\imath\Delta}{3}\sqrt{\frac{2}{3}} & \frac{(1-\imath)\Delta}{3\sqrt{2}} & 0 & -\frac{(1+\imath)\Delta}{3\sqrt{6}}\\
-\frac{(1+\imath)\Delta}{3\sqrt{6}} & -\frac{\imath\Delta}{3}\sqrt{\frac{2}{3}} & \frac{\lambda}{2} & \frac{(1+\imath)\Delta}{3\sqrt{3}}& \frac{\imath\Delta}{3\sqrt{3}} & 0 \\
0 & \frac{(1+\imath)\Delta}{3\sqrt{2}} & \frac{(1-\imath)\Delta}{3\sqrt{3}} & \frac{\lambda}{2} & 0 & \frac{\imath\Delta}{3\sqrt{3}} \\
\frac{(1-\imath)\Delta}{3\sqrt{2}} & 0 & -\frac{\imath\Delta}{3\sqrt{3}} & 0 & \frac{\lambda}{2} & -\frac{(1+\imath)\Delta}{3\sqrt{3}}  \\
-\frac{\imath\Delta}{3}\sqrt{\frac{2}{3}} & -\frac{(1-\imath)\Delta}{3\sqrt{6}} & 0 & -\frac{\imath\Delta}{3\sqrt{3}} & -\frac{(1-\imath)\Delta}{3\sqrt{3}} & \frac{\lambda}{2}
\end{array} \right ).
\end{eqnarray}
\end{widetext}
Diagonalization of ${\hat H}$ leads to three doublets at energies
$$E^{(1,2)}=-\frac{\Delta}{6} -\frac{\lambda }{4}-\frac{1}{2}\sqrt{2\lambda ^{2}+(\Delta -\frac{\lambda }{2})^{2}},$$ corresponding to eigenstates $|\Phi_1\rangle$ and $|\Phi_2\rangle$,
$$E^{(3,4)}=-\frac{\Delta}{6} -\frac{\lambda }{4}+\frac{1}{2}\sqrt{2\lambda ^{2}+(\Delta -\frac{\lambda }{2})^{2}},$$ corresponding to eigenstates $|\Phi_3\rangle$ and $|\Phi_4\rangle$, and $$E^{(5,6)}=\frac{\Delta}{3} +\frac{\lambda }{2},$$ corresponding to eigenstates $|\Phi_5\rangle$ and $|\Phi_6\rangle$. Within the ground state doublet ($|\Phi_1\rangle$ and $|\Phi_2\rangle$) we choose the orthonormal states such that the $J_{\rm eff}^z=\pm1/2$ states do not mix with each other as  mentioned above.

\subsection{Two-hole states}\label{secIIB}

In paper I,\cite{perkins14} we  explained  how to obtain two-hole eigenstates. We refer the reader to this paper for details, as we only briefly outline  the main steps and set notations here.

The full two-hole Hamiltonian is the sum of two contributions: a single-particle term, $H_{\lambda,\Delta}$, which includes the SO coupling and trigonal CF, and the many-body part, $H_{\rm int}$,
 given by the Coulomb interaction, $U_2$,  and the Hund’s coupling, $J_H$ (Eq. (6) in paper I).  There are
 $6\times 5/2=15$ partly degenerate two-hole eigenstates obtained by diagonalization of the full on-site Hamiltonian
 \begin{eqnarray}\label{hamfull}
 H_{{\rm int}+\lambda,\Delta}\equiv H_{\rm int}+H_{\lambda,\Delta}~.
 \end{eqnarray}
 We denote energy eigenstates of the full Hamiltonian  (\ref{hamfull}) as
\begin{eqnarray}
\vert D,\xi \rangle =\sum_{\mu =1}^{15}c_{\xi \mu
}\vert {\mathcal \Phi\mathcal\Phi},\mu \rangle~,
\end{eqnarray}
where the  two-hole basis states $\vert {\mathcal \Phi\mathcal\Phi},\mu \rangle$ are  simply given by  direct products of eigenstates $|\Phi_1\rangle, ... |\Phi_6\rangle$ diagonalizing one-particle Hamiltonian (\ref{HAM}):
\begin{eqnarray}\label{phiphi}
\begin{array}{lll}
\vert {\mathcal \Phi\mathcal\Phi},1\rangle &\equiv&
\vert { \Phi}_1{\Phi}_2\rangle\\[0.1cm]
\vert {\mathcal \Phi\mathcal\Phi},2\rangle &\equiv &
\vert {\Phi}_1{\Phi}_3\rangle\\[0.1cm]
\vert {\mathcal \Phi\mathcal\Phi},3\rangle &\equiv &
\vert { \Phi}_1{\Phi}_4\rangle\\[0.1cm]
\vert {\mathcal \Phi\mathcal\Phi},4\rangle &\equiv &
\vert { \Phi}_1{\Phi}_5\rangle\\[0.1cm]
\vert {\mathcal \Phi\mathcal\Phi},5\rangle &\equiv &
\vert { \Phi}_1{\Phi}_6\rangle\\[0.1cm]
\vert {\mathcal \Phi\mathcal\Phi},6\rangle &\equiv &
\vert { \Phi}_2{\Phi}_3\rangle\\[0.1cm]
\vert {\mathcal \Phi\mathcal\Phi},7\rangle &\equiv &
\vert { \Phi}_2{\Phi}_4\rangle\\[0.1cm]
\vert {\mathcal \Phi\mathcal\Phi},8\rangle &\equiv &
\vert { \Phi}_2{\Phi}_5\rangle\\[0.1cm]
\vert {\mathcal \Phi\mathcal\Phi},9\rangle &\equiv &
\vert { \Phi}_2{\Phi}_6\rangle\\[0.1cm]
\vert {\mathcal \Phi\mathcal\Phi},10\rangle &\equiv &
\vert { \Phi}_3{\Phi}_4\rangle\\[0.1cm]
\vert {\mathcal \Phi\mathcal\Phi},11\rangle &\equiv &
\vert { \Phi}_3{\Phi}_5\rangle\\[0.1cm]
\vert {\mathcal \Phi\mathcal\Phi},12\rangle &\equiv &
\vert { \Phi}_3{\Phi}_6\rangle\\[0.1cm]
\vert {\mathcal \Phi\mathcal\Phi},13\rangle &\equiv &
\vert { \Phi}_4{\Phi}_5\rangle\\[0.1cm]
\vert {\mathcal \Phi\mathcal\Phi},14\rangle &\equiv &
\vert { \Phi}_4{\Phi}_6\rangle\\[0.1cm]
\vert {\mathcal \Phi\mathcal\Phi},15\rangle &\equiv &
\vert
{ \Phi}_5{\Phi}_6\rangle
\end{array}
\end{eqnarray}
We denote by
$c_{\xi \mu }$ and $E_{\xi }$, correspondingly,  the eigenvectors and eigenvalues and $\xi =1,...15$.

\section{Derivation of the super-exchange Hamiltonian}\label{sec:ham}

 The super-exchange process which couples the magnetic moments of Ir$^{4+}$ ions originating from the Kramers' doublet ground states involves intermediate states
 with either zero holes or two holes.
 As discussed in Sec.\ref{secIIB}, the latter states are governed by the Coulomb  and the Hund's interaction, as well as  by the  SO coupling and the trigonal CF.
The connection between the Kramers' doublet ground states $\Phi_1$ and $\Phi_2$  at site $n$ ($\gamma =1,2$) and the full manifold of $\Phi$-states at site $n^{\prime}$ ($\gamma^{\prime} =1,2,...,6$)
is  given by the projected hopping term:
\begin{eqnarray}
PH_{t,n,n^{\prime}}=\sum_{\gamma =1}^{2}\sum_{\gamma ^{\prime
}=1}^{6}T_{n,n^{\prime}}^{\gamma ,\gamma ^{\prime }}b_{n,\gamma }^\dagger b_{n^{\prime},\gamma
^{\prime }}~,
\end{eqnarray}
where the elements of the matrix $T_{n,n^{\prime}}^{\gamma ,\gamma ^{\prime }}$  will be derived in the next section.
For the moment, let us derive the super-exchange Hamiltonian treating $T_{n,n^{\prime}}^{\gamma ,\gamma ^{\prime }}$
 as generic hopping matrix between  either n. n.  or next n. n. Ir$^{4+}$ ions.

The super-exchange
Hamiltonian, obtained by  the second order perturbation  theory,  can be written as
\begin{eqnarray}\label{hamex-formal}
H_{{\rm ex},n,n^{\prime }}=\sum_{\xi }\frac{1}{\epsilon _{\xi }}PH_{t,n,n^{\prime
}}Q_{\xi ,n^{\prime }}H_{t,n^{\prime },n}P~,
\end{eqnarray}
where
\begin{eqnarray}
P=\prod_n\sum_{\sigma
_{n}=\pm 1}\vert 1/2,\sigma _{n}/2;n\rangle \langle
n;1/2,\sigma_{n} /2\vert
\end{eqnarray}
is the projection operator onto the  ground states with one hole at site $n$. The projection
operators onto two-hole intermediate states $\vert D,\xi;n^{\prime} \rangle $
with excitation energy $\epsilon_{\xi }$ at site $n^{\prime}$ are given by
 \begin{eqnarray}
 Q_{\xi ,n^{\prime}}=\vert D,\xi ;n^{\prime}\rangle \langle n^{\prime};D,\xi
\vert= D_{\xi ,n^{\prime}}^\dagger D_{\xi ,n^{\prime}}~.
\end{eqnarray}
 The excitation energies
of the intermediate states are  $\epsilon _{\xi }=E_{0h}+E_{\xi }-2E_{1h}$.
Rewriting operator $D_{\xi ,n}$ as  $D_{\xi ,n}=\sum_{\nu =1}^{15}\sum_{\gamma _{1},\gamma
_{2}=1}^{6}c_{\xi ,\nu }m_{\gamma _{1}\gamma _{2}}^{\nu }b_{\gamma
_{1},n}^{\dagger}b_{\gamma _{2},n}^{\dagger}$, where by
$b_{\gamma,n}^\dagger $ we denote an operator creating a hole  of the type $\gamma=1,...6$, which  refers to  the component of the single-hole vector ${\hat \Phi}$  at the site $n$ and
the tensor ${\hat m}$ has
only two non-zero elements for each state $\nu$:
\begin{eqnarray}\nonumber
m_{1,2}^{1}&=&m_{1,3}^{2}=m_{1,4}^{3}=m_{1,5}^{4}
=m_{1,6}^{5}=\\\nonumber
m_{2,3}^{6}&=&m_{2,4}^{7}=m_{2,5}^{8}=m_{2,6}^{9}
=m_{3,4}^{10}=\\\nonumber
m_{3,5}^{11}&=&m_{3,6}^{12}=m_{4,5}^{13}=m_{4,6}^{14}=m_{5,6}^{15}=1
\end{eqnarray}
and
\begin{eqnarray}\nonumber
m_{2,1}^{1}&=&m_{3,1}^{2}=m_{4,1}^{3}=m_{5,1}^{4}
=m_{6,1}^{5}=\\\nonumber
m_{3,2}^{6}&=&m_{4,2}^{7}=m_{5,2}^{8}=m_{6,2}^{9}
=m_{4,3}^{10}=\\\nonumber
m_{5,3}^{11}&=&m_{6,3}^{12}=m_{5,4}^{13}=m_{6,4}^{14}=m_{6,5}^{15}=-1~.
\end{eqnarray}

It is convenient to rewrite the Hamiltonian (\ref{hamex-formal}) in the  second-quantized form:
\begin{eqnarray}\label{hamex}
H_{{\rm ex},n,n^{\prime}}&=&\sum_{\sigma ,\sigma ^{\prime }=1}^{2}\sum_{\sigma
_{1},\sigma _{1}^{\prime }=1}^{2}\sum_{\xi =1 }^{15}\\\nonumber
&&\frac{1}{\epsilon _{\xi }}%
\{A_{n,n^{\prime};\sigma ,\sigma ^{\prime }}^{\xi }b_{n,\sigma }^\dagger b_{n^{\prime},\sigma
^{\prime }}^\dagger A_{n^{\prime},n;\sigma _{1}^{\prime },\sigma _{1}}^{\xi
}b_{n^{\prime},\sigma _{1}^{\prime }}b_{n,\sigma _{1}}\}~,
\end{eqnarray}
where  we have defined coefficients $A_{n,n^{\prime};\sigma ,\sigma ^{\prime }}^{\xi }$ as
\begin{eqnarray}\label{def:A}
A_{n,n^{\prime};\sigma ,\sigma ^{\prime }}^{\xi }=\sum_{\gamma _{1}=1}^{6}
\sum_{\nu =1}^{15}T_{n,n^{\prime}}^{\sigma ,\gamma _{1}}c_{\xi ,\nu
}(m_{\gamma _{1}\sigma ^{\prime }}^{\nu }-m_{\sigma ^{\prime }\gamma
_{1}}^{\nu })~.
\end{eqnarray}

Next, we define the magnetic degrees of freedom with the help of
the pseudospin operators
$S_{n}^{\alpha }=\frac{1}{2}\sum_{\sigma
,\sigma ^{\prime }=\pm 1}\tau _{\sigma ,\sigma ^{\prime }}^{\alpha
}b_{\sigma ,n}^\dagger b_{\sigma ^{\prime },n}$  and the density operator
$\rho _{n}=\sum_{\sigma =\pm 1}b_{\sigma ,n}^\dagger b_{\sigma ,n}$.  With $\alpha =x,y,z$,  we denote the spin component index
and $\tau _{\sigma ,\sigma ^{\prime }}^{\alpha}$ are the Pauli matrices.   Then,  the super-exchange Hamiltonian  (\ref{hamex}) on the bond $n,n^{\prime}$  can be written in  terms
of the magnetic  degrees of freedom of Ir$^{4+}$ as
\begin{eqnarray}\label{spinham}
H_{{\rm ex},n,n^{\prime}}=\sum_{\alpha\beta} \Gamma^{\alpha\beta }_{n,n^{\prime}}S_n^{\alpha}S_{n^{\prime}}^{\beta}+
W \rho
_{n}\rho _{n^{\prime}},
\end{eqnarray}
$\alpha,\beta $ label Cartesian components of pseudospins.
The first term represents the most general bilinear form of the super-exchange Hamiltonian. The second term gives a constant energy shift and we shall hereafter omit it. We also note that because of time reversal symmetry, there are no terms of  the kind $S_n^{\alpha}\rho _{n^{\prime}}$.
The exchange coupling matrix $\Gamma^{\alpha\beta }$ on the bond $n,n^{\prime}$ has the form
\begin{eqnarray} \label{gamma}
\Gamma_{n,n^{\prime}}=
\left( \begin{array}{ccc}
J^x\,\,&J^{xy}\,\,&J^{xz}\\
J^{yx}\,\,&J^{y}\,\,&J^{yz}\\
J^{zx}\,\,&J^{zy}\,\,&J^{z}\\
\end{array} \right)~
\end{eqnarray}
and its elements are  given in the Appendix. In the following, we shall call  $\Gamma_1^{\alpha\beta }$ and $\Gamma_2^{\alpha\beta }$  the exchange coupling matrix for nearest and second nearest neighbors, respectively.
 Because of the lack of the tight-binding parameters for third nearest neighbors, we
 will not derive the  $\Gamma_3^{\alpha\beta }$  matrix and treat  the third neighbor coupling as isotropic.

\section{The hopping matrix}\label{sec:hop}

\subsection{The nearest neighbors hopping matrix}

 In A$_2$BO$_3$ compounds, the honeycomb lattice of Ir$^{4+}$ ions is embedded in the cubic lattice and corresponds to  one of the (111) planes. Three  kinds of honeycomb lattice bonds, denoted as $x,\,y$ and $z$ and drawn by red, green and blue solid lines in Fig. 1 (a), correspond
 to the cubic face diagonals along vectors (0,1,1), (1,0,1) and (1,1,0), respectively.

We first consider the hopping matrix between neighboring Ir$^{4+}$  ions. The strongest  n. n.  hopping is via an intermediate oxygen ion. For each pair of  n. n. Ir$^{4+}$  ions, there are two  Ir-O-Ir paths and the total hopping amplitude arises as a sum of these two hoppings. The direct hopping between nearest Ir ions is also not negligible due to the extended nature of $5d$ orbitals. Thus, the total hopping Hamiltonian  comes from two contributions: $H_t=H_{\rm O-assist}+H_{\rm dir}$.

We focus our discussion on the hopping along  a single $z$-bond because the system is translationally invariant and contributions from  $x$ and $y$ bonds can be obtained by rotational symmetry.
 Along the $z-$bond, the 90$^{\circ}$ hopping  occurs via  $p_{z}-$orbitals of oxygen ions, which, following
  Ref.\cite{jackeli09},  we call  the upper and the lower one (see Fig. 1 (b)). The upper $p_{z}-$orbital overlaps
with the $ X$ orbital of the Ir$^{4+}$ ion on the A sublattice and with the $Y$ orbital  on the B sublattice.
  	Vice versa, the lower $p_{z}-$orbital overlaps
with the $ Y$ orbital of the Ir$^{4+}$ ion on the A sublattice and with the $X$ orbital of the Ir$^{4+}$ ion on the B sublattice.
The overlaps of  $ X$ and $p_{z}$ and $ Y$ and $p_{z}$ are equal. Thus, we have $t_{X,z}=t_{Y,z}=t_{pd\pi}$.
 We  next integrate out the upper oxygen ion  and compute the effective hopping between Ir$^{4+}$ ions  through the upper Ir-O-Ir bond. The amplitude of the effective Ir-Ir hopping  is then equal to $t_{1o}=t_{pd\pi}^2/\Delta_p$ and $\Delta_p$ stands for the charge transfer gap.
  The hopping via the lower oxygen is just the complex conjugate of the hopping via the upper oxygen.
  The direct hopping along  a  $z$-bond has the biggest matrix element for   diagonal hopping between nearest $Z$ orbitals. We denote the amplitude of this hopping as $t_d$. In our calculations  for n. n. hoppings, we will use the value of the oxygen assisted hopping equal to $t_{1o}= 230$ meV and the  direct hopping equal to $t_d=67$ meV.
  These values were
  obtained by  Foyevtsova {\it et al}.~\cite{katerina13} by tight-binding fitting of ab-initio electronic structure calculations in the presence of trigonal distortion.

For the ultimate derivation of the super-exchange Hamiltonian we do not need the whole $6\times6$ hopping matrix  but
only its first two lines connecting ground state doublet $ \Phi_1$ and $\Phi_2$ to all six states belonging to
${\hat \Phi}$. Combining contributions from the  two paths (via the upper and via the lower oxygens), and adding direct hopping, we obtain the effective hopping
Hamiltonian between n. n.  Ir$^{4+}$  ions along the $z$-bond
\begin{eqnarray}
H_{t}^z=\sum_{n}\sum_{\gamma ,\gamma ^{\prime }}T_{1,n,n+z}^{\gamma ,\gamma
^{\prime }}(b_{n,\gamma }^\dagger b_{n+z,\gamma ^{\prime }}+h.c.),
\end{eqnarray}
where $b_{\gamma,n}^\dagger$ is an operator creating a hole on site $n$ of the type $\gamma=1,...6$, which  refers to  the components of the vector ${\hat \Phi}$.
The hopping matrix is given by
\begin{widetext}
\begin{eqnarray}\label{Hopping matrix}
T_{1,n,n+z}=
{\small\left( \begin{array}{cccccc}
\langle \Phi_1\vert {\hat T_1}\vert \Phi_1\rangle & \langle \Phi_1\vert {\hat T_1}\vert \Phi_2\rangle & \langle \Phi_1\vert {\hat T_1} \vert \Phi_3\rangle&\langle \Phi_1\vert {\hat T_1}\vert \Phi_4\rangle&\langle \Phi_1\vert {\hat T_1} \vert\Phi_5\rangle&\langle \Phi_1\vert {\hat T_1} \vert\Phi_6\rangle\\
\langle \Phi_2\vert {\hat T_1} \vert\Phi_1\rangle & \langle \Phi_2\vert {\hat T_1}\vert \Phi_2 \rangle& \langle \Phi_2\vert {\hat T_1} \vert \Phi_3\rangle&\langle \Phi_2\vert {\hat T_1}\vert \Phi_4\rangle&\langle \Phi_2\vert {\hat T_1}\vert \Phi_5\rangle&\langle \Phi_2\vert {\hat T_1}\vert \Phi_6\rangle
\end{array} \right)}
\end{eqnarray}
\end{widetext}

Let us analyze the structure of the hopping matrix (\ref{Hopping matrix}) in the absence of trigonal distortion, $\Delta=0$. In this case, the single-hole vector ${\hat \Phi}$ is nothing else but the vector $
{\hat J}=\{
|\frac{1}{2},\frac{1}{2}\rangle,
|\frac{1}{2},-\frac{1}{2}\rangle,
|\frac{3}{2},\frac{3}{2}\rangle,
|\frac{3}{2},\frac{1}{2}\rangle ,
|\frac{3}{2},-\frac{1}{2}\rangle,
|\frac{3}{2},-\frac{3}{2}\rangle\}
$ diagonalizing the SO interaction. In this limit,
the two transfer amplitudes via upper
and lower oxygen interfere in a destructive manner and, because of this,
the  only non-zero elements of the effective transfer matrix are $$T_{n,n+z}^{1 ,6}=T_{n,n+z}^{2 ,3}=-\frac{2\imath}{\sqrt{6}}t_{1o}^2$$ and their complex conjugates, where $\gamma=1,2$ correspond to $\vert 1/2,\pm 1/2\rangle$ and $\gamma=3,6$  correspond to $\vert 3/2,\pm 3/2\rangle$ states.
  As was shown  by Jackeli and Khaliullin,\cite{jackeli09}
    this massive cancelation of hopping terms in the absence of trigonal distortion leads to a vanishing  isotropic part of the super-exchange  mediated by   oxygen ions. The  non-zero  n. n. isotropic  term is, therefore, entirely determined by the direct hopping $t_d$  between $d$-orbitals of the Ir ions.

\begin{figure*}
\label{fig2}
\includegraphics[width=0.95\columnwidth]{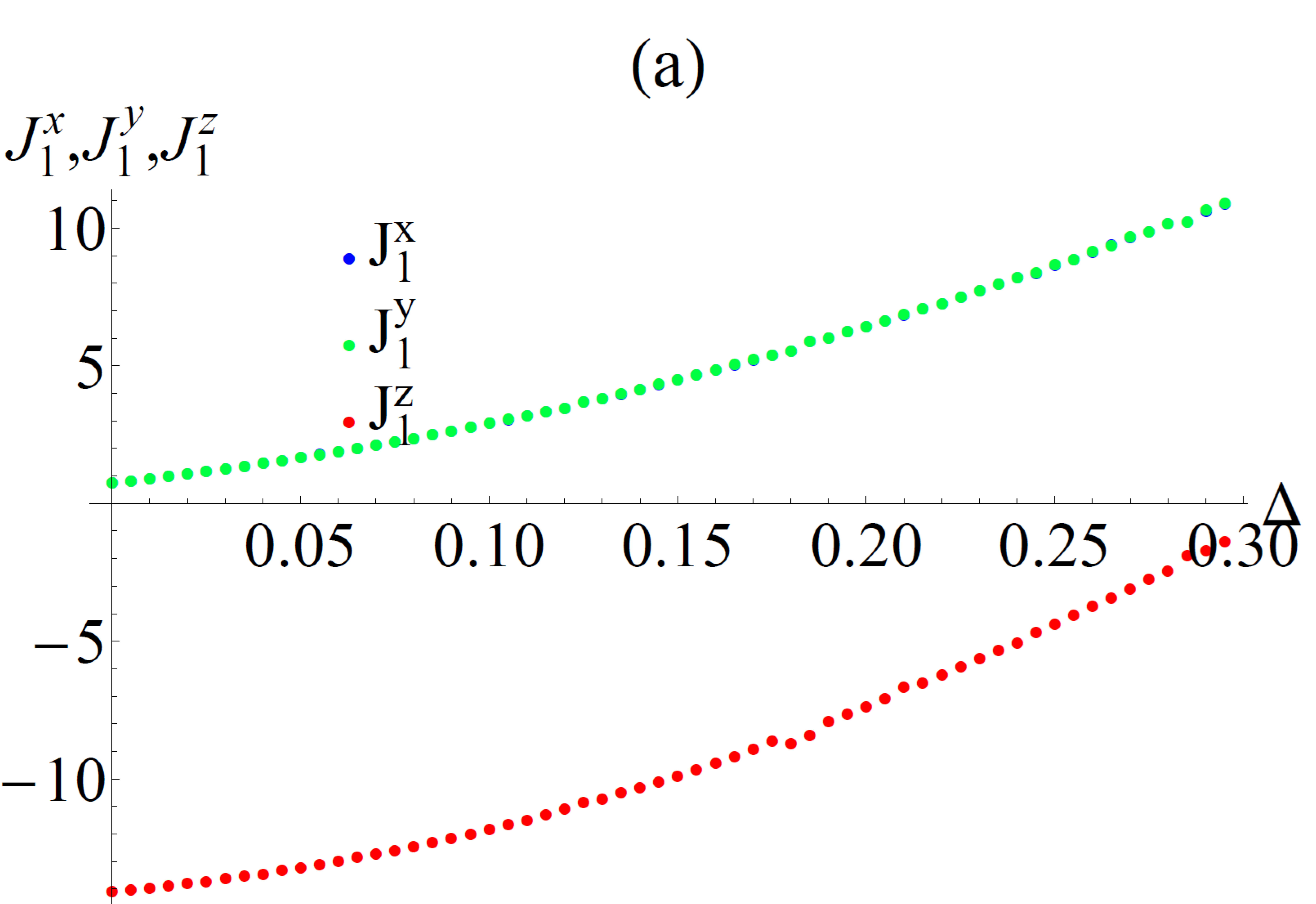}
\includegraphics[width=0.95\columnwidth]{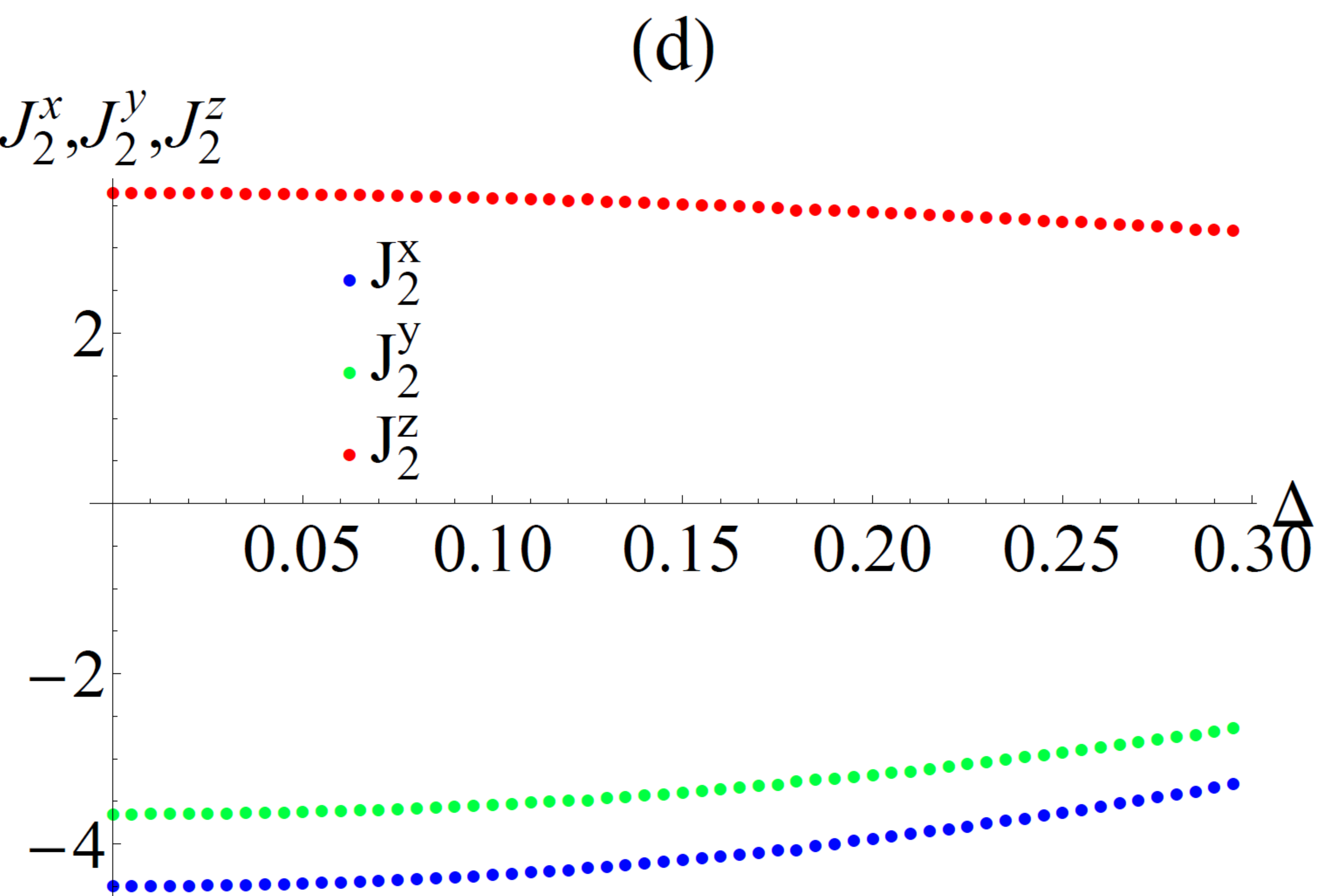}
\includegraphics[width=0.95\columnwidth]{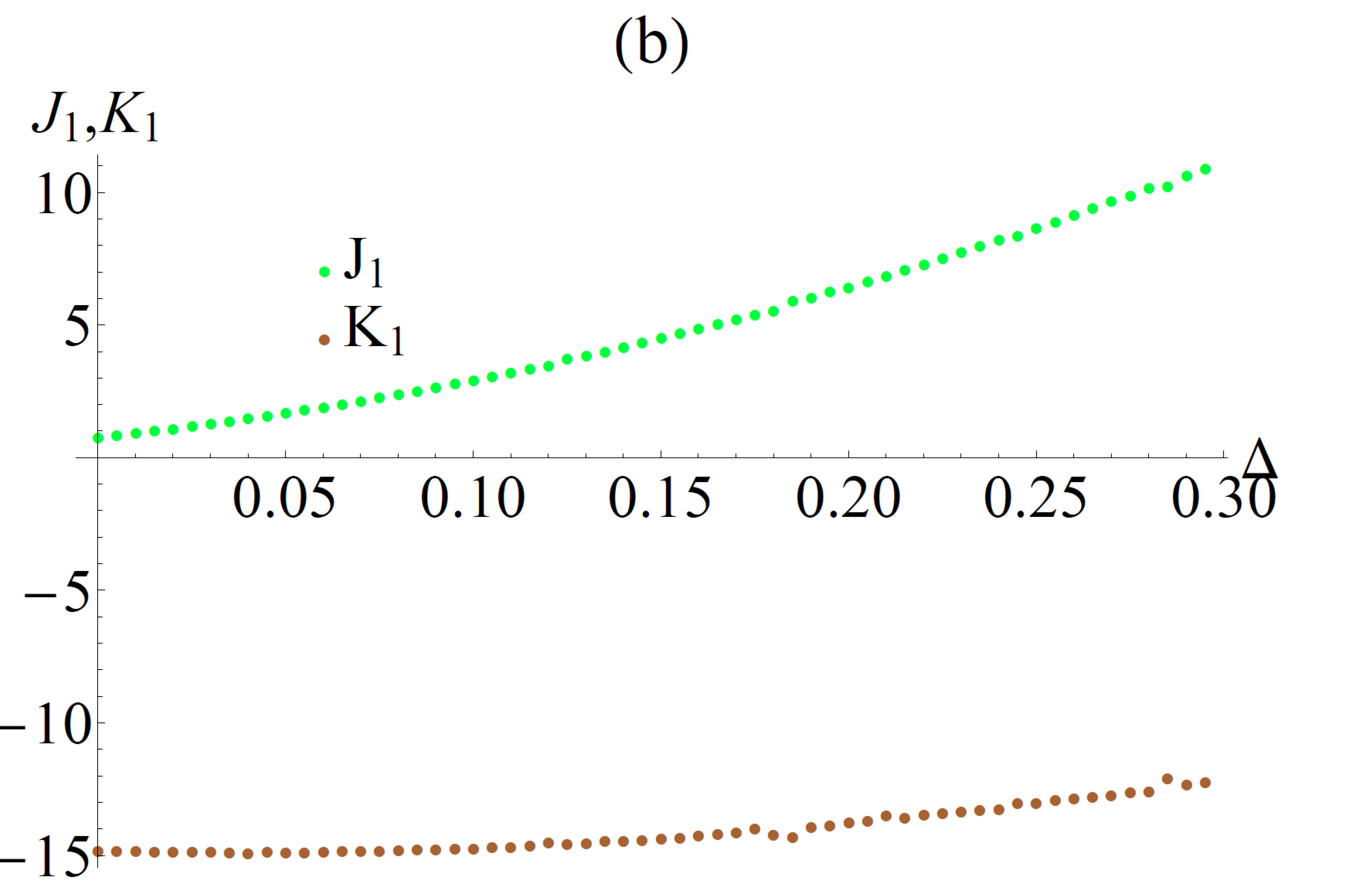}
\includegraphics[width=0.95\columnwidth]{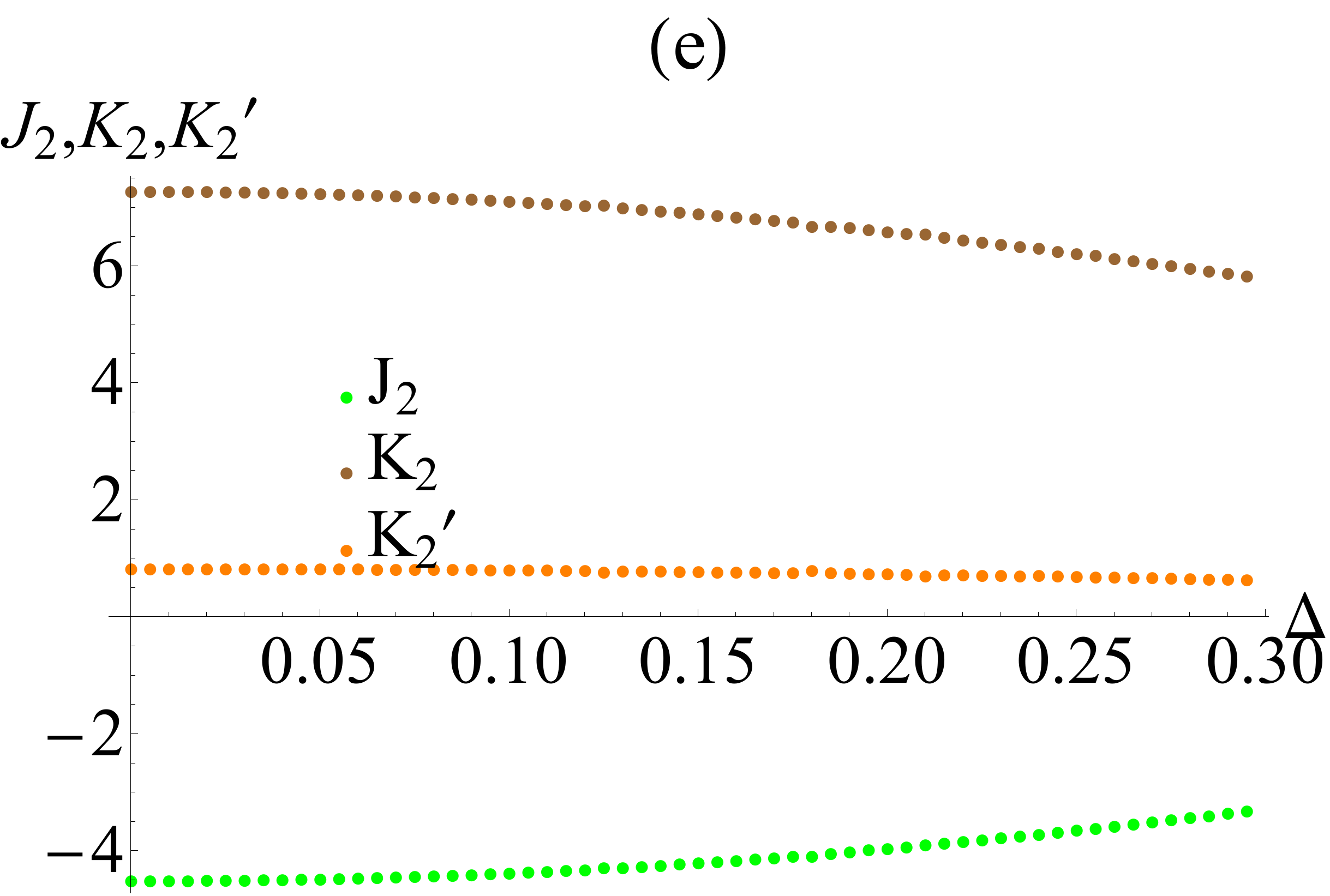}
\includegraphics[width=0.95\columnwidth]{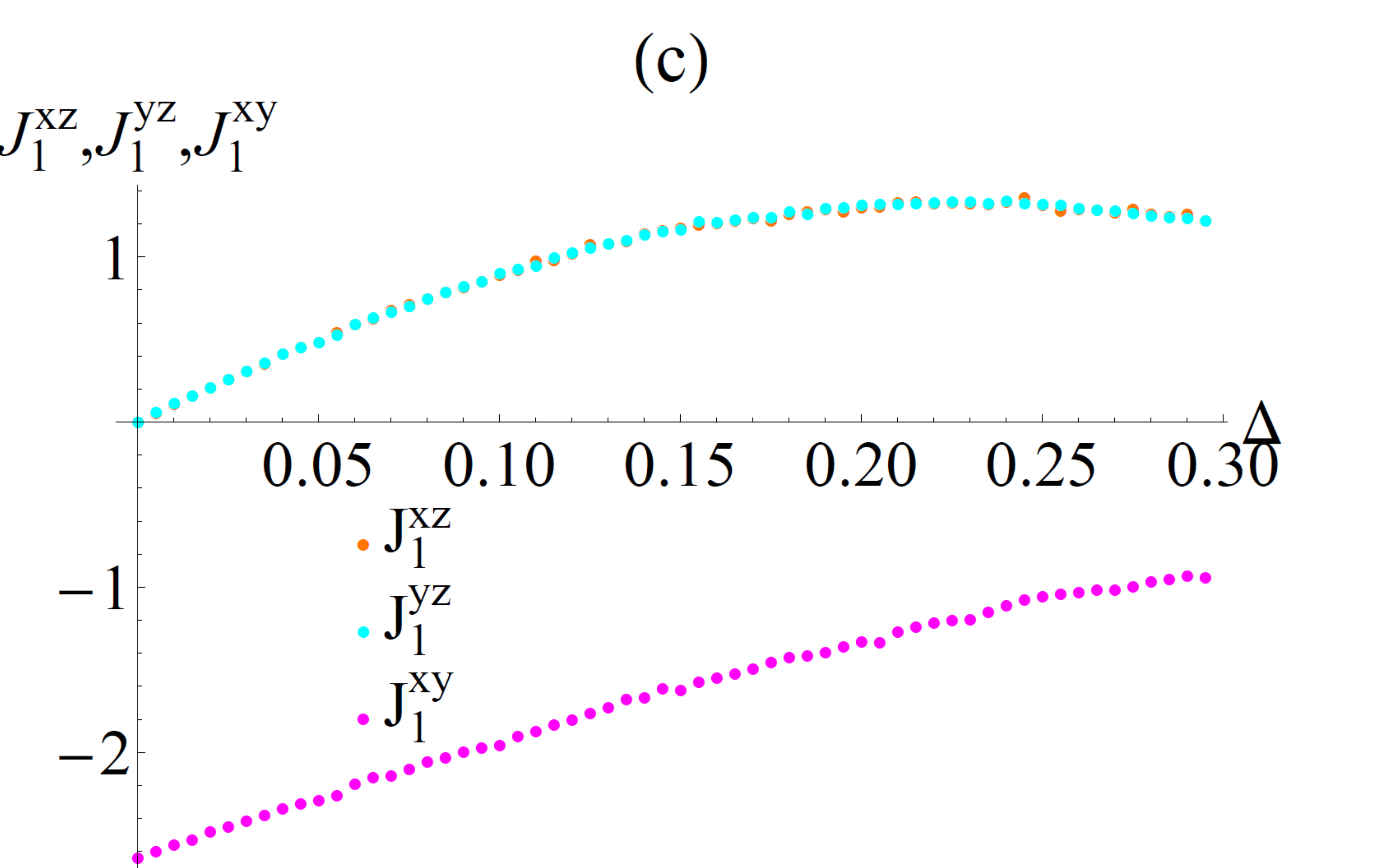}
\includegraphics[width=0.95\columnwidth]{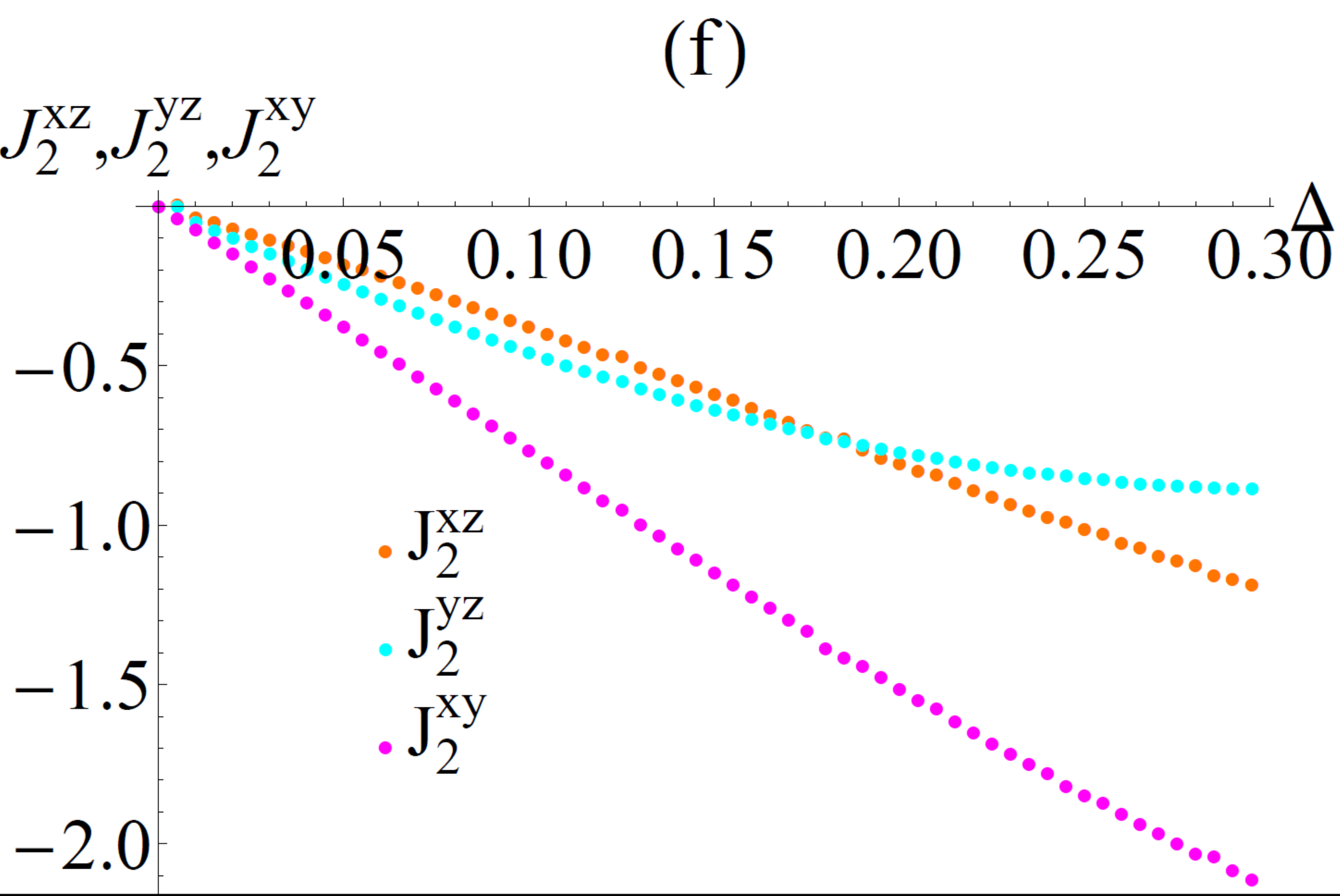}
\caption{(Colors online)
(a) The diagonal exchange couplings $J_{1}^x,J_{1}^y,J_{1}^z$  and (d) $J_{2}^x,\,J_{2}^y,\,J_{2}^z$ in meV (shown by blue, green  and red lines, respectively); (b) the n. n. Kitaev interaction $K_1$ and the n. n. isotropic exchange $J_1$   and (e) the second neighbor Kitaev interactions $K_2$ and  $K_2^\prime$, as well as the second neighbor isotropic exchange $J_2$   in meV (shown by brown, orange and green lines, respectively); (c) the off-diagonal exchange couplings $ J_1^{xy},\, J_1^{xz}, \,J_1^{yz}$  and (f)
$ J_2^{xy}, \,J_2^{xz},\,J_2^{yz}$ in meV (shown by magenta, orange, cyan lines, respectively)
plotted as function of trigonal  crystal field $\Delta$ (in eV).
 The  microscopic parameters of the model are considered to be $J_H=0.3$ eV, $U_2=1.8$ eV,
 $\lambda=0.4$ eV, $t_{1o}=230$ meV, $t_{d}=67$ meV and  $t_{2o}=95$ meV.
 }
\end{figure*}

\subsection{The second neighbor hopping matrix}

Next, we derive the hopping matrix for second neighbors.
Six  bonds between second neighbors Ir$^{4+}$ ions on the honeycomb lattice correspond
 to (2,1,-1), (1,2,1),  (-1,1,2), (-2,-1,1), (-1-2,-1), (1,-1,-2) bonds, which we call ${\tilde x}$, ${\tilde y}$, ${\tilde z}$, and ${\tilde x}$, ${\tilde y}$, ${\tilde z}$ bonds, respectively. Then, the second neighbor ${\tilde x}-$bond connects two Ir ions which are also connected by   two n. n.  Ir-Ir bonds of $y-$ and $z-$ type,  and ${\tilde y}-$ and  ${\tilde z}-$bonds connect Ir$^{4+}$ ions which are connected by  $x-$ and $z-$, and $x-$ and $y-$bonds, respectively. In Fig. 1 (a), we also use the same color coding for the second neighbor bonds as for n. n. bonds: ${\tilde x}-$, ${\tilde y}-$, ${\tilde z}-$ bonds are shown by red, green and blue dotted lines.

Similarly to the hopping between  nearest neighbors, there are also two kinds of hoppings  connecting second neighbors (see Fig. 1 (a)): the hopping along the path Ir-O-Na-O-Ir, and  the direct one.
The indirect hopping $t_{2o}$ is  large both because   it comes from four Ir-O-Na-O-Ir paths  but also because it  takes advantage of the extended nature of the $s-$orbital of the Na ion.
In the ideal structure, it is equal to $t_{2o} =82.1$ meV, and in the presence of the trigonal distortion it is even larger, $t_{2o} =94.7$ meV.\cite{katerina13}  The direct hopping between second neighbors is significantly smaller than the one between nearest neighbors and also significantly smaller than the hopping along the Ir-O-Na-O-Ir path.  In our derivation of the  second neighbor super-exchange Hamiltonian, we will neglect all second neighbor hoppings except  $t_{2o}$.

Explicitly, the hopping matrix element between second neighbor Ir ions along the ${\tilde z}$-bond  comes from the following processes:\cite{comment}
\begin{eqnarray*}
&\text{Path 1}:\,
{\rm Ir}\,(Y)\rightarrow
{\rm O}\,(p_z)\rightarrow
{\rm Na}\,(s)\rightarrow
{\rm O}\,(p_z)\rightarrow
{\rm Ir}\,(X)\\
&\text{Path 2}:\,
{\rm Ir}\,(Y)\rightarrow
{\rm O}\,(p_z)\rightarrow
{\rm Na}\,(s)\rightarrow
{\rm O}\,(p_y)\rightarrow
{\rm Ir}\,(X)\\
&\text{Path 3}:\,
{\rm Ir}\,(Y)\rightarrow
{\rm O}\,(p_x)\rightarrow
{\rm Na}\,(s)\rightarrow
{\rm O}\,(p_z)\rightarrow
{\rm Ir}\,(X)\\
&\text{Path 4}:\,
{\rm Ir}\,(Y)\rightarrow
{\rm O}\,(p_x)\rightarrow
{\rm Na}\,(s)\rightarrow
{\rm O}\,(p_y)\rightarrow
{\rm Ir}\,(X)\\
\end{eqnarray*}

Summing over all these four  paths, shown by thick magenta lines in Fig. 1 (a), we obtain
the effective hopping
Hamiltonian between second neighbor Ir$^{4+}$ ions along the ${\tilde z}$-bond
\begin{eqnarray}
H_{t}^{\tilde z}=\sum_{n}\sum_{\gamma ,\gamma ^{\prime }}T_{2,n,n+{\tilde z}}^{\gamma ,\gamma
^{\prime }}(b_{n,\gamma }^\dagger b_{n+{\tilde z},\gamma ^{\prime }}+h.c.),
\end{eqnarray}
where, formally, the hopping matrix $T_{2,n,n+{\tilde z}}$ has the same structure as  $T_{1,n,n+{ z}}$ given
by Eq. (\ref{Hopping matrix}).
\begin{figure*}
\label{fig3}
\includegraphics[width=0.95\columnwidth]{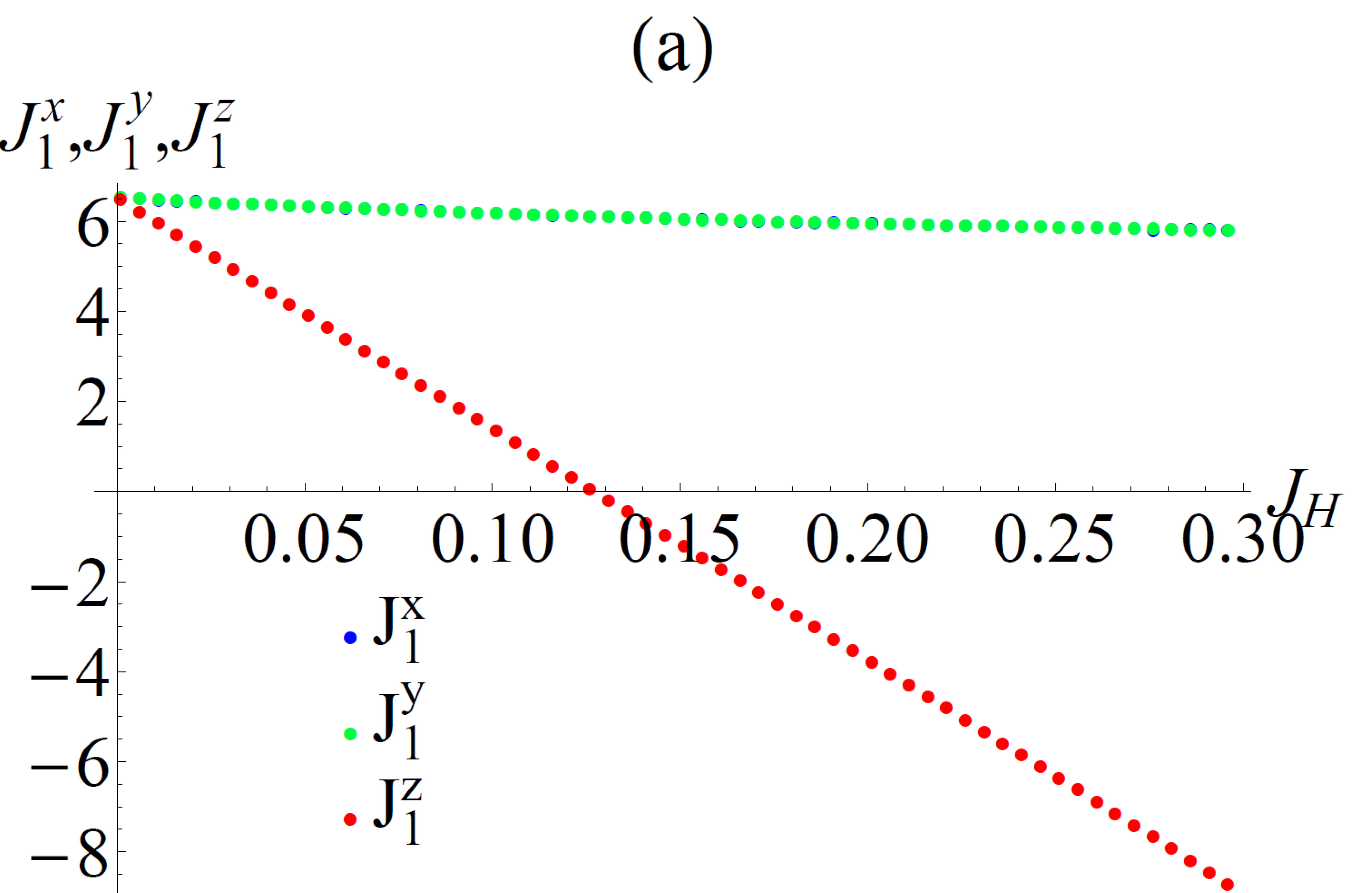}
\includegraphics[width=0.95\columnwidth]{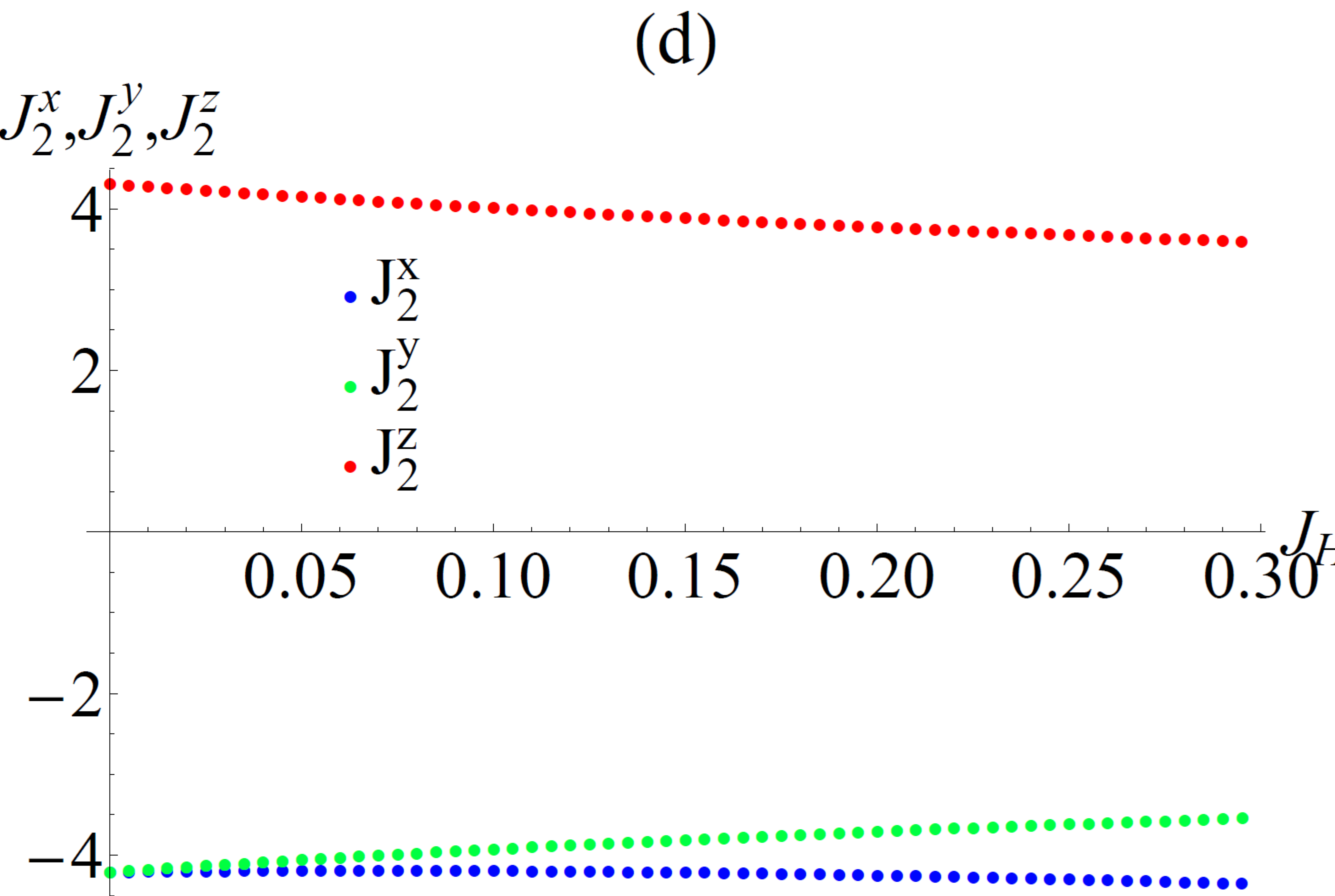}
\includegraphics[width=0.95\columnwidth]{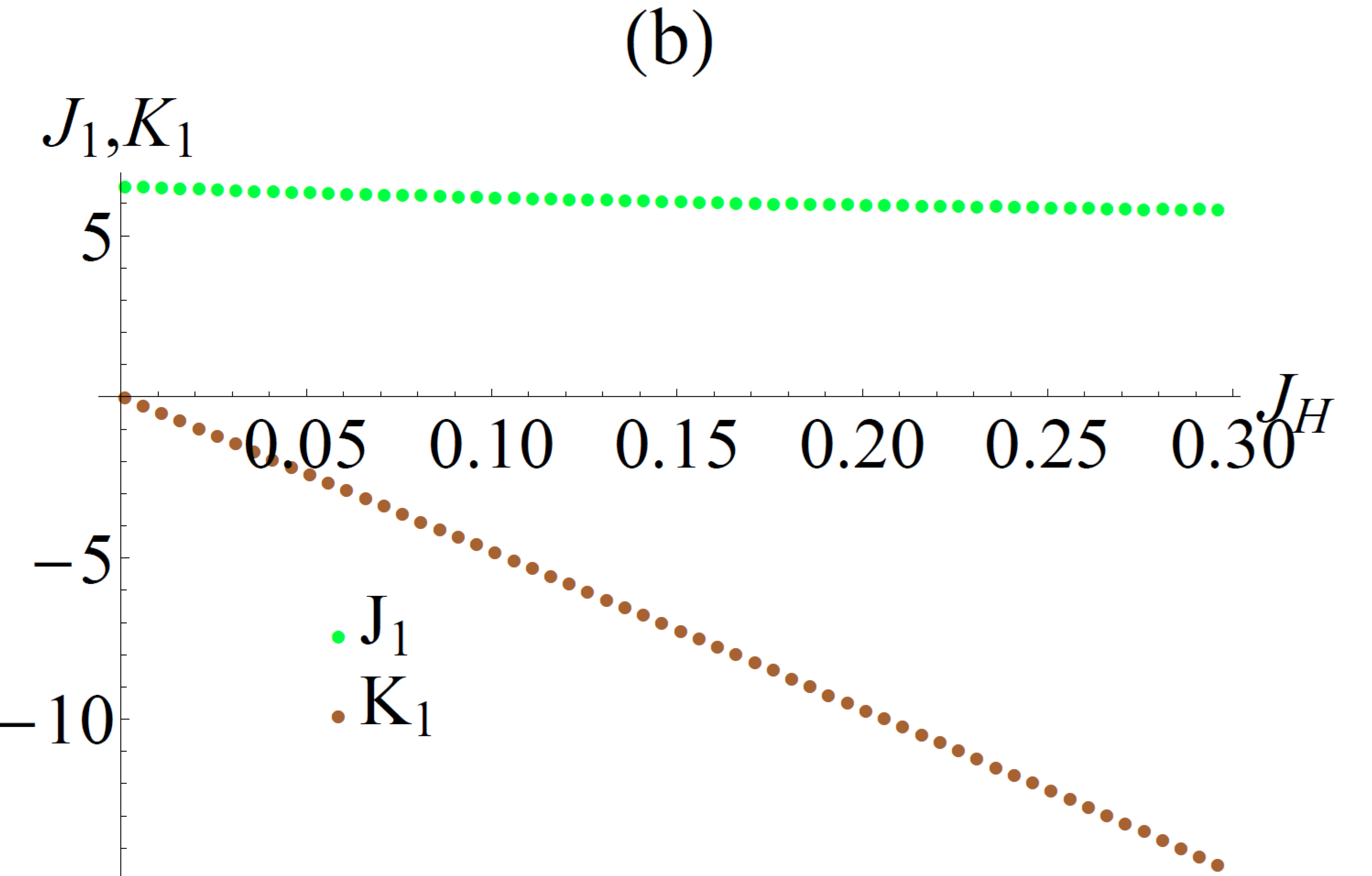}
\includegraphics[width=0.95\columnwidth]{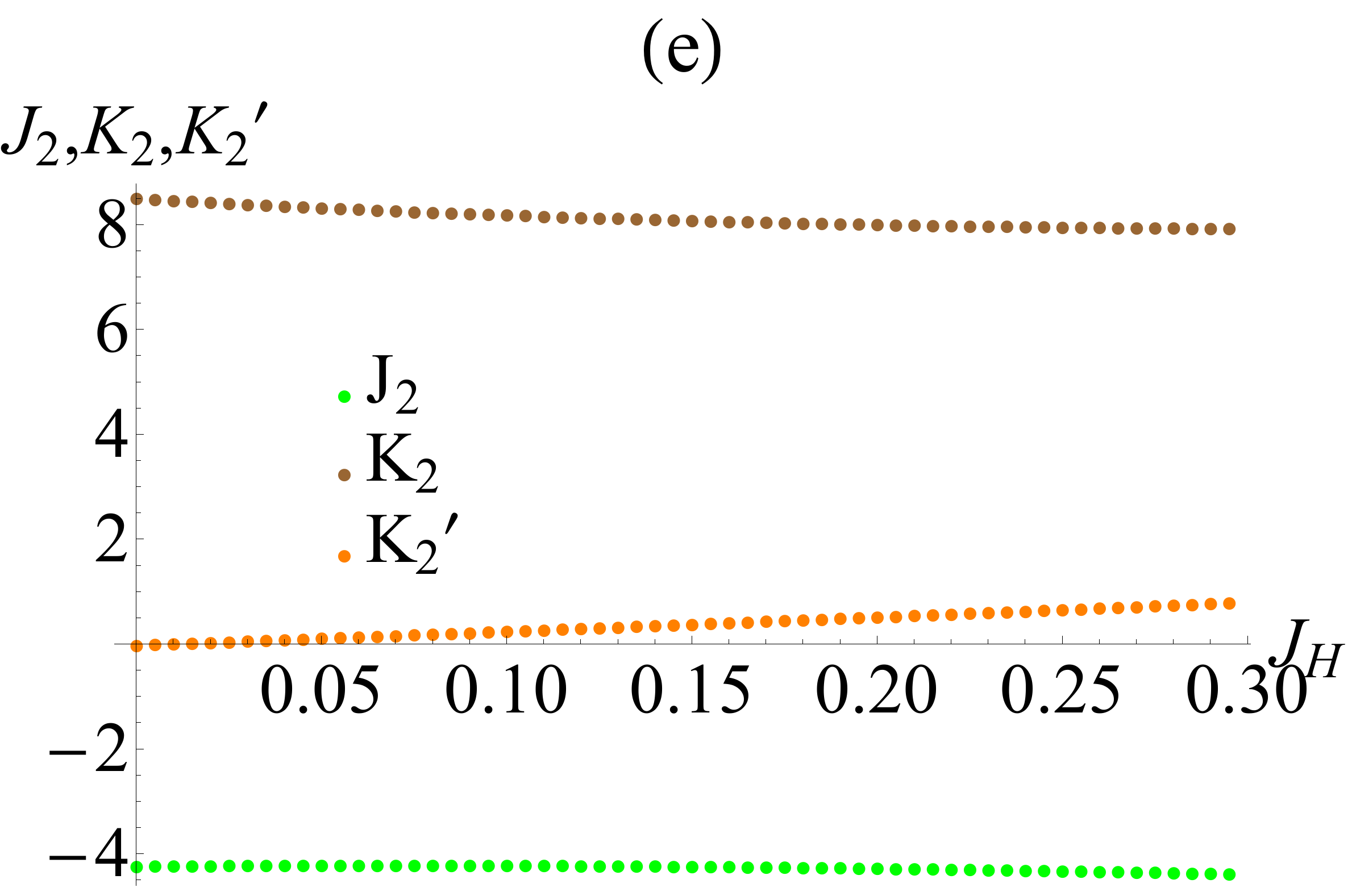}
\includegraphics[width=0.95\columnwidth]{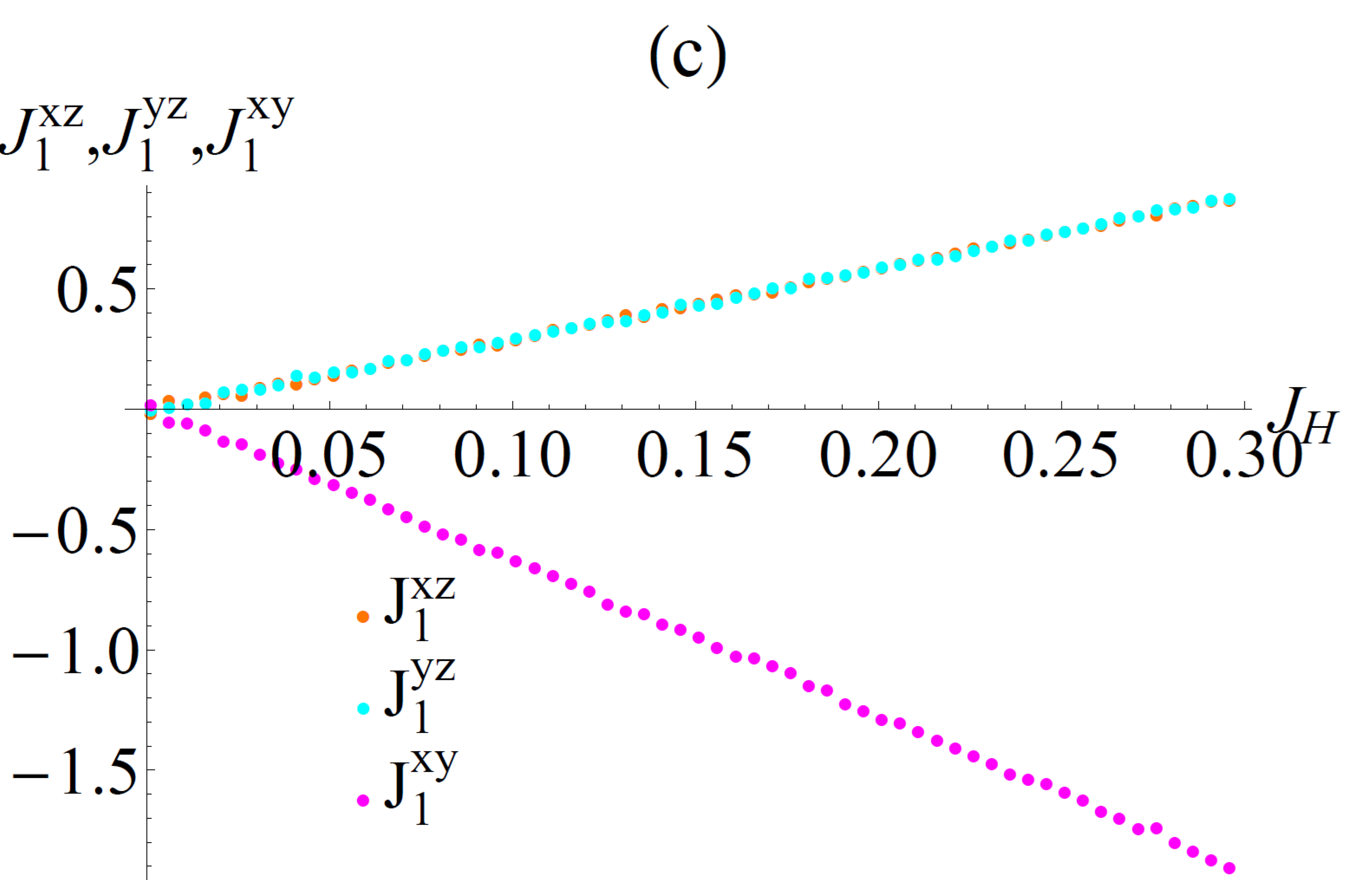}
\includegraphics[width=0.95\columnwidth]{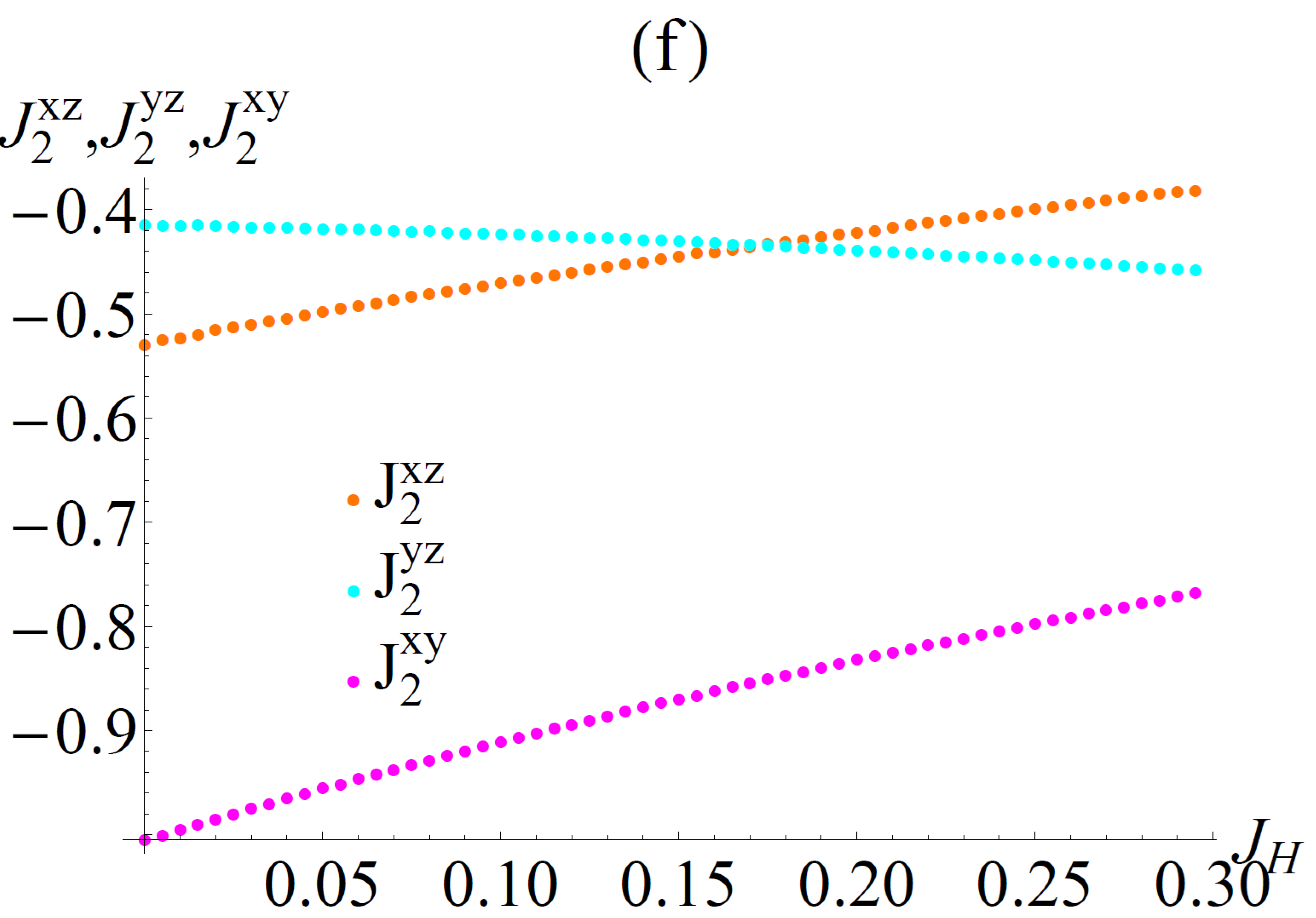}
\caption{(Colors online)
(a) The diagonal exchange couplings $J_{1}^x,\,J_{1}^y,\,J_{1}^z$  and (d) $J_{2}^x,\,J_{2}^y,\,J_{2}^z$ in meV (shown by  blue, green, red  lines, respectively);  (b) the n. n. Kitaev interaction $K_1$ and the n. n. isotropic exchange $J_1$   and (e) the second neighbor Kitaev interactions $K_2$ and  $K_2^\prime$, as well as the second neighbor isotropic exchange $J_2$   in meV (shown by brown, orange and green lines, respectively);
(c) the off-diagonal exchange couplings $ J_1^{xy},\, J_1^{xz},\,J_1^{yz}$  and (f)
$ J_2^{xy},\, J_2^{xz},\,J_2^{yz}$ in meV (shown by magenta, orange, cyan, magenta lines, respectively)
 plotted as
as functions of Hund's coupling, $J_H$ (in eV).
 The  microscopic parameters of the model are considered to be $U_2=1.8$ eV,
 $\lambda=0.4$ eV, $\Delta=0.1$ eV, $t_{1o}=230$ meV, $t_{d}=67$ meV and  $t_{2o}=95$ meV.
}
\end{figure*}

\section{The  exchange coupling tensors $\Gamma_1^{\alpha\beta }$ and $\Gamma_2^{\alpha\beta }$}\label{sec:coupling}

 We show in Fig. 2 and Fig. 3  how the matrix elements of the  exchange coupling tensor $\Gamma^{\alpha\beta }$,
defined in Eq.(\ref{gamma}), computed for both nearest  and second neighbor Ir$^{4+}$ ions  depend on the microscopic parameters
(trigonal distortion, Hund's coupling, Coulomb interaction and SO coupling). We note right away that the main role of the Coulomb repulsion is to determine the overall energy scale for  the couplings. Thus, in all computations we  take, for definitiveness, $U_2=1.8$ eV, which is laying inside the range of values, 1.5 eV-2.5 eV,  characteristic  to iridates. We  also set the SO coupling constant to  $\lambda=0.4$ eV since it is the value  associated with Ir$^{4+}$ ions in the literature. As we already mentioned before, we compute all exchange interactions for either $z$-nearest  or for ${\tilde z}$  next n. n. bonds. Interactions for other bonds can be obtained using symmetry arguments.

\subsection{Effect of trigonal distortion.}

 Here we study the dependencies of the exchange couplings on the trigonal distortion, $\Delta$. 
At ambient pressure, the trigonal crystal field splitting in both Na$_2$IrO$_3$ and Li$_2$IrO$_3$ is about 110 meV.\cite{gret13}
 However, it is also believed that a much  stronger trigonal distortion can be reached under pressure.
In this subsection, the exchange parameters were computed  for a fixed Hund's coupling, $J_H=0.3$ eV.

In Fig. 2 (a)-(c),  we  plot  the $\Delta$-dependencies  of   the  matrix elements of the tensor $\Gamma_1^{\alpha\beta }$ on the $z$-bond.
 In order to define n. n. Kitaev interactions on $x$- and $y$-bonds, one needs to permute indices of bonds and couplings which is done with the help of Fig. 4. 
 The diagonal matrix elements $J_{1}^x,\,J_{1}^y,\,J_{1}^z$ are shown in Fig. 2 (a).
We see that while the $J_{1}^x$ and $J_{1}^y$ couplings  are positive and  degenerate for all values of the  trigonal splitting, $J_{1}^x=J_{1}^y=J_1$, the  $J_{1}^z$ coupling is first negative but then changes sign at $\Delta\simeq 0.2$ eV.
 The  anisotropic n. n. Kitaev interaction,  $K_1$,  may be defined as  the difference between diagonal elements. On the $z$-bond,  it is simply given by $K_1\equiv J_{1}^z-J_{1}$.
   We plot  $J_1$ and $K_1$
  in Fig. 2 (b).  Notice that  while the n. n. isotropic  exchange is {\it antiferromagnetic}  and  is rapidly growing with $\Delta$, the  Kitaev interaction is {\it ferromagnetic} and is almost independent of the magnitude of the trigonal field.

 In Fig. 2 (d)-(f),  we  plot  the $\Delta$-dependencies  of   the  matrix elements
    of the tensor $\Gamma_2^{\alpha\beta }$. We see that the second neighbor diagonal elements $J_{2}^x,\,J_{2}^y,\,J_{2}^z$, presented in Fig. 2 (d), are substantially weaker than the n. n. diagonal interactions (see Fig. 2 (a)).
  There is  also no degeneracy between them: all of the second neighbor diagonal elements are different from each other except $ J_{2}^z=-J_{2}^y$.   If we define the isotropic exchange as $ J_{2}^y=J_2$, and anisotropic second neighbor Kitaev interactions as  $K_2\equiv J^z_{2}-J^y_{2}=-2J_2$ and $K_2^\prime\equiv J^y_{2}-J^x_{2}$,
      then the interaction on the ${\tilde z}$-bond can be written as
$J_2 {\mathbf S}{\mathbf S}+K_2 S^zS^z- K_2^\prime S^xS^x$.   We plot $J_2,\,K_2$ and $K_2^{\prime}$ as a function of $\Delta$ in Fig. 2 (e). Note that for all values of $\Delta$  $J_2<0$, $K_2>0$ and $K_2^{\prime}>0$, and also $K_2\gg K_2^{\prime}$.

It is also important to remember that $J_2$, $K_2$ and $K_2^\prime$ all come from the same process and  are governed by the same hopping parameter $t_{2o}$. 
This is in the contrast to  the n. n. couplings, $J_1$  and $K_1$,  for which the super-exchange processes in the absence of the trigonal distortion are completely distinct.
$J_1$ is  determined by the direct hopping, with  amplitude $t_{d}$, and $K_1$ is determined with amplitude $t_{1o}$, mediated by the hopping through the intermediate oxygen.
 The interactions between second neighbors come from the same process and  are governed by the same hopping parameter $t_{2o}$.

 The behavior of the off-diagonal terms $ J_1^{xy}, \,J_1^{xz}, \,J_1^{yz}$ and  $ J_2^{xy},\, J_2^{xz},\,J_2^{yz}$  is shown in Fig. 2 (c) and (f), respectively. At $\Delta=0$, all of them, except $J_1^{xy}$, are equal to zero. The non-zero value of $J_1^{xy}$ is due to the finite value of the Hund's coupling. As we  will see in the next subsection, $J_1^{xy}(J_H=0)=0$. The magnitudes of all off-diagonal terms grow with the strength of the trigonal distortion, however they remain subdominant interactions even at relatively large $\Delta$.

\begin{figure}
\label{fig4}
\includegraphics[width=0.95\columnwidth]{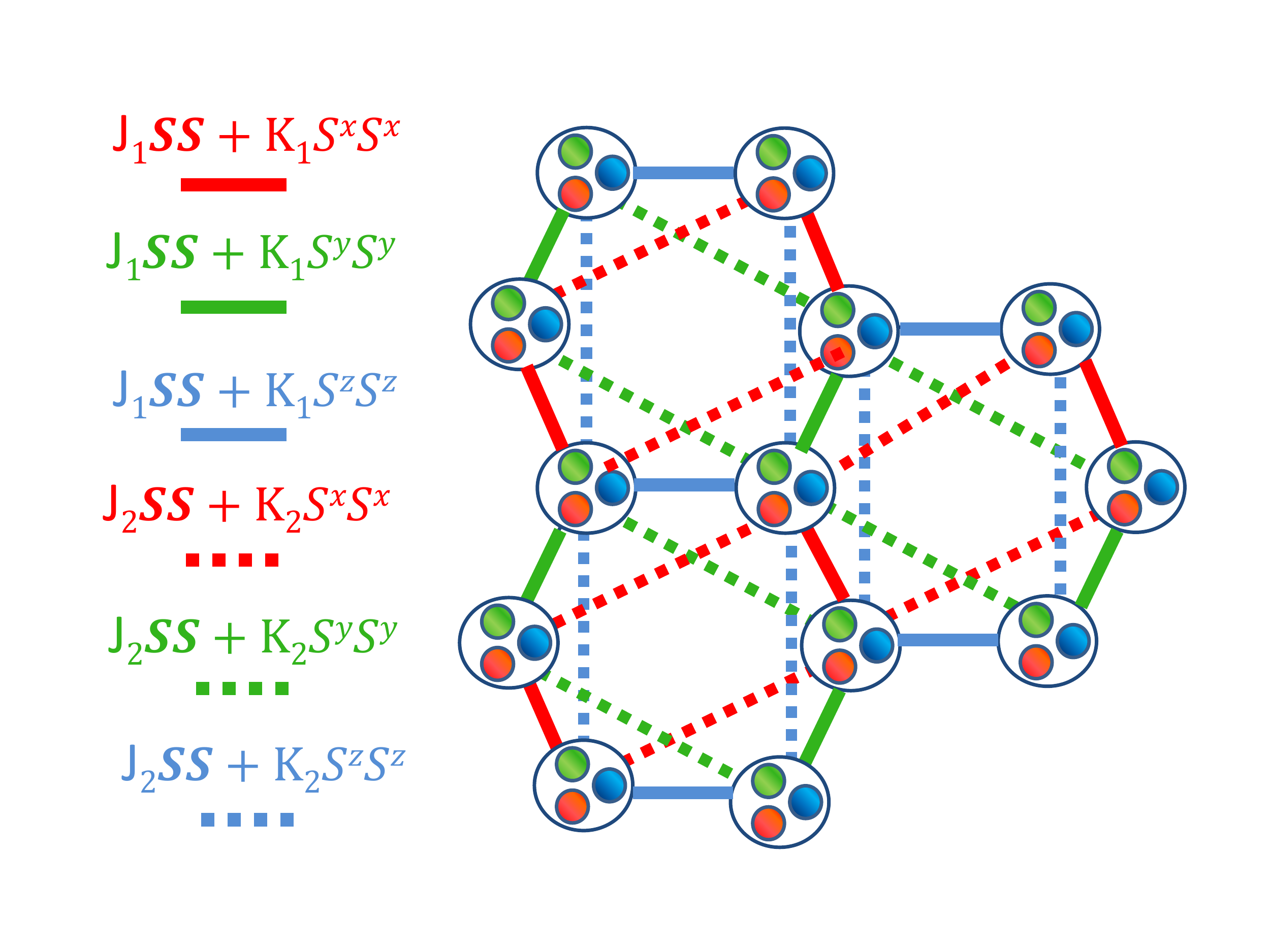}
\caption{(Colors online)  Schematic representation of the effective  super-exchange model for Na$_2$IrO$_3$. Color coding is the same as in Fig. 1 (a).  $X$, $Y$ and $Z$ $t_{2g}$ electronic orbitals, participating in the super-exchange, are shown by red, green and blue small circles.
}
\end{figure}

\subsection{Effect of Hund's coupling.}

In Fig. 3, we present the dependence of the exchange couplings on the Hund's interaction, $J_H$. Here we fix  the trigonal distortion equal to $\Delta=0.1$ eV.

In  Fig. 3 (a) and (d), we plot $J_{1}^x,\,J_{1}^y,\,J_{1}^z$ and $J_{2}^x,\,J_{2}^y,\,J_{2}^z$, respectively.
   At $J_H=0$, we see that  the n. n. diagonal couplings are all equal, $J_{1}^x=J_{1}^y=J_{1}^z$.  Consequently,  the n. n. Kitaev interaction is $K_1=0$.  The n. n. off-diagonal couplings (see Fig. 3 (c)) are also zero at $J_H=0$.
   On the contrary,
   the next n. n.  diagonal couplings are only partially degenerate: $J_{2}^x=J_{2}^y=- J_{2}^z$.  Thus,  $K_2\neq 0$ and $K_2^\prime \neq 0$.
   The second neighbor off-diagonal couplings (see Fig. 3 (f)) are all non-zero but very small. 
    Thus, at $J_H=0$ the leading anisotropic term is the Kitaev interaction between second neighbors,  $K_2$.
    With increasing $J_H$,  the n. n. Kitaev interaction, $K_1$,  rapidly grows and, at realistic values of Hund's coupling, about 0.2-0.3 eV, becomes the dominant interaction. 
    With increasing $J_H$,  $K_1$ rapidly grows and  becomes the dominant interaction at realistic values of Hund's coupling, about 0.2-0.3 eV.
    The other  exchange couplings also change with $J_H$. Overall,   the n. n. interactions are more sensitive to the strength of the Hund's coupling than the second neighbors.

Let us summarize the results obtained in this section. The most important anisotropies resulting from our microscopic calculations are the Kitaev interactions on n. n. and next n. n. bonds, $K_1$ and $K_2$ respectively. All other anisotropic interactions remain subdominant for reasonable values of miscroscopic parameters. $K_1$ is weakly dependent on the trigonal CF, but grows quickly with Hund's coupling. However, $K_2$ depends weakly on both $\Delta$ and $J_H$.

\section{Magnetic phase diagram}\label{sec:pd}

\subsection{Effective super-exchange model for Na$_2$IrO$_3$.}

We now discuss how  the above results  apply to the case of Na$_2$IrO$_3$.
We take the values of the microscopic parameters most closely related to Na$_2$IrO$_3$: $\lambda=0.4$ eV, $\Delta=0.1$ eV, $J_H=0.3$ eV,  $U_2=1.8$ eV, and hopping matrix elements  equal to $t_{1o}=230$ meV, $t_{d}=67$ meV and  $t_{2o}=95$ meV.\cite{katerina13}
We obtain the following  exchange couplings:
$J_1=5.8$ meV, $K_1=-14.8$ meV, $J_2=-4.4$ meV, $K_2=7.9$ meV.
Calculated n. n. exchange constants are in fair agreement with the results of ab-initio quantum chemistry calculations by Katukuri {\it et al}:\cite{katukuri14} $J_1\simeq 3$ meV and $K_1\simeq -17.5$ meV.

 Our results  for  n. n. couplings
confirm  the previous conclusion\cite{singh10,singh12,liu11,ye12,choi12} that the super-exchange model with only n. n. couplings  is insufficient to  explain the  experimentally observed zigzag magnetic order even in the presence of  the trigonal distortion. Recall that in the original  Kitaev-Heisenberg model,\cite{jackeli09,jackeli10}
  the  isotropic and Kitaev exchange couplings were  parameterized by a single parameter $\alpha$ as
  $J_1=1-\alpha$  and $K_1=2\alpha$. Taking $J_1$  and $K_1$  obtained for the trigonal distortion $\Delta\simeq$0.1 eV,  we get $\alpha\simeq 0.57$, which corresponds to the stripy antiferromagnetic order instead of the zigzag-type order. Neglecting the trigonal distortion and taking $J_1=1.4$ meV and $K_1=-15.2$ meV
 obtained at $\Delta=$0 eV,  we get  $\alpha\simeq 0.83$ corresponding to the spin liquid, which was desired but not observed in  Na$_2$IrO$_3$.\cite{singh10,singh12,liu11,ye12,choi12}

This shows that, in addition to the  {\it antiferromagnetic}  Heisenberg and {\it ferromagnetic} Kitaev n.n. interactions, the minimal model has to include further neighbor interactions.
As we saw in Sec.\ref{sec:coupling}, the dominant microscopic  Ir-Ir couplings also include  next n. n. {\it ferromagnetic} Heisenberg and {\it antiferromagnetic} Kitaev interactions, which also must be considered.

Thus,  let  us study the  following super-exchange Hamiltonian:
%\begin{eqnarray}\label{minimal}
$$\mathcal{H}=J_1\sum_{\langle n,n^{\prime}\rangle_{\gamma}}{\bf S}_n{\bf S}_{n^{\prime}}
+K_1\sum_{\langle n,n^{\prime}\rangle_{\gamma}} S_n^{\gamma}S_{n^{\prime}}^{\gamma}
$$%\nonumber\\
\begin{equation}\label{minimal}
+J_2\sum_{\langle\langle n,n^{\prime}\rangle\rangle_{\tilde\gamma}}
{\bf S}_n{\bf S}_{n^{\prime}}
+K_2 \sum_{\langle\langle n,n^{\prime}\rangle\rangle_{\tilde\gamma}}S_n^{\gamma}S_{n^{\prime}}^{\tilde\gamma}
\end{equation}
%\\\nonumber
$$+J_3\sum_{\langle\langle\langle n,n^{\prime}\rangle\rangle\rangle}
{\bf S}_n{\bf S}_{n^{\prime}},
%\end{equation}
$$
where $J_1>0$, $K_1<0$, $J_2<0$, $K_2=-2J_2>0$, and $J_3>0$.
 Note that in our formulation of the minimal model (\ref{minimal}), we also include  the third neighbor antiferromagnetic coupling,  which was suggested to be crucial for stabilizing the zigzag magnetic order in the previous  works.\cite{kimchi11,choi12}

It is very important  that  the presence of the second  n. n.  Kitaev interaction  does not change the space group symmetries of the effective model: the model (\ref{minimal}) has the same symmetries  as the original Kitaev-Heisenberg model.
The schematic representation of the n.n. and  second n. n. interactions is shown in Fig. 4. As in Fig. 1 (a), the solid lines correspond to n. n. bonds and dotted lines correspond to the second  n. n.  Kitaev interaction.
 We also note that the same form of the second neighbor interactions was  previously obtained\cite{rachel10,reuther12}
 in the limit $U\rightarrow\infty$ of the Kane-Mele-Hubbard model.\cite{kane05}

\begin{figure*}
\label{fig5}
\includegraphics[width=1.0\columnwidth]{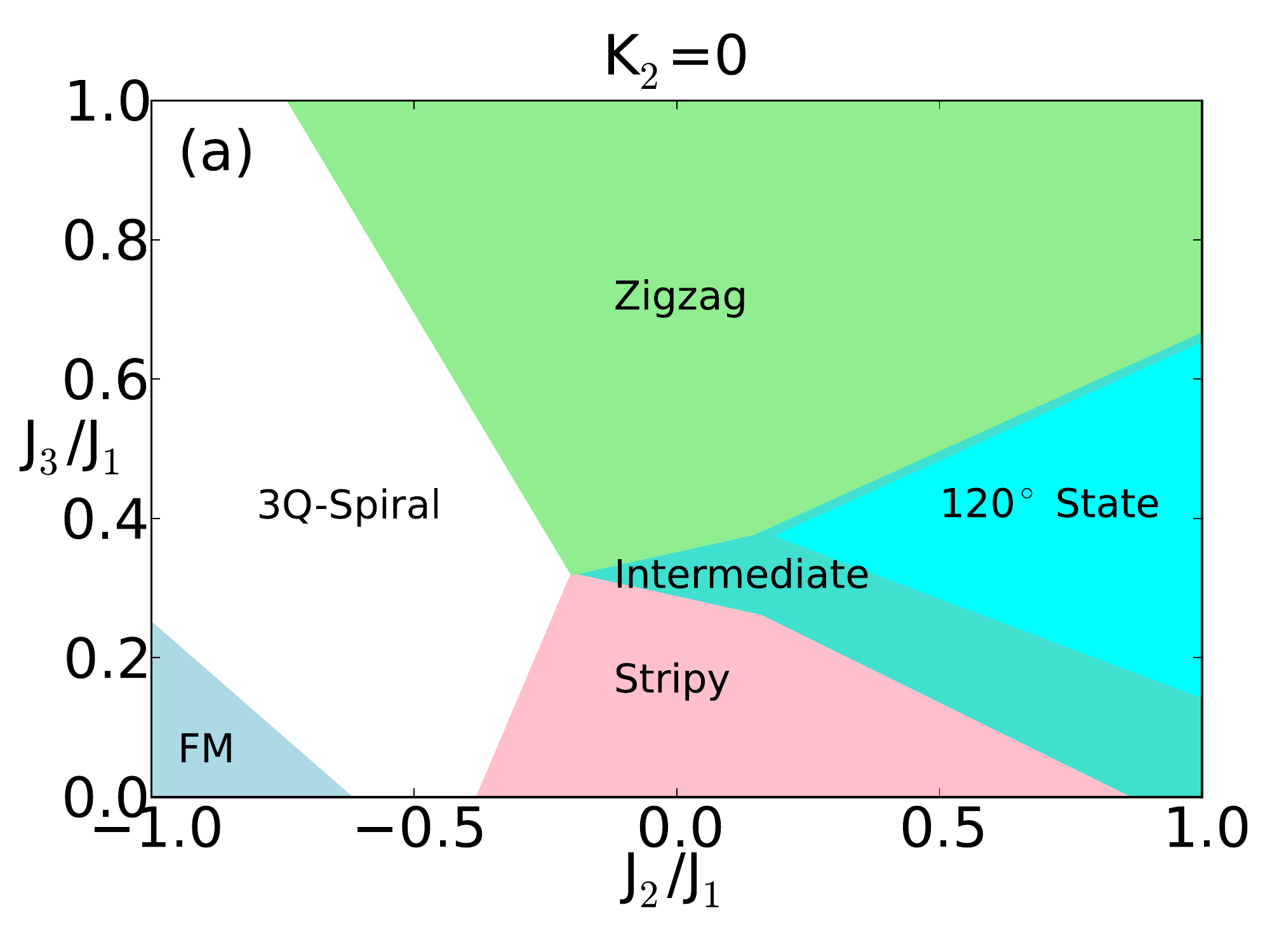}
\includegraphics[width=1.0\columnwidth]{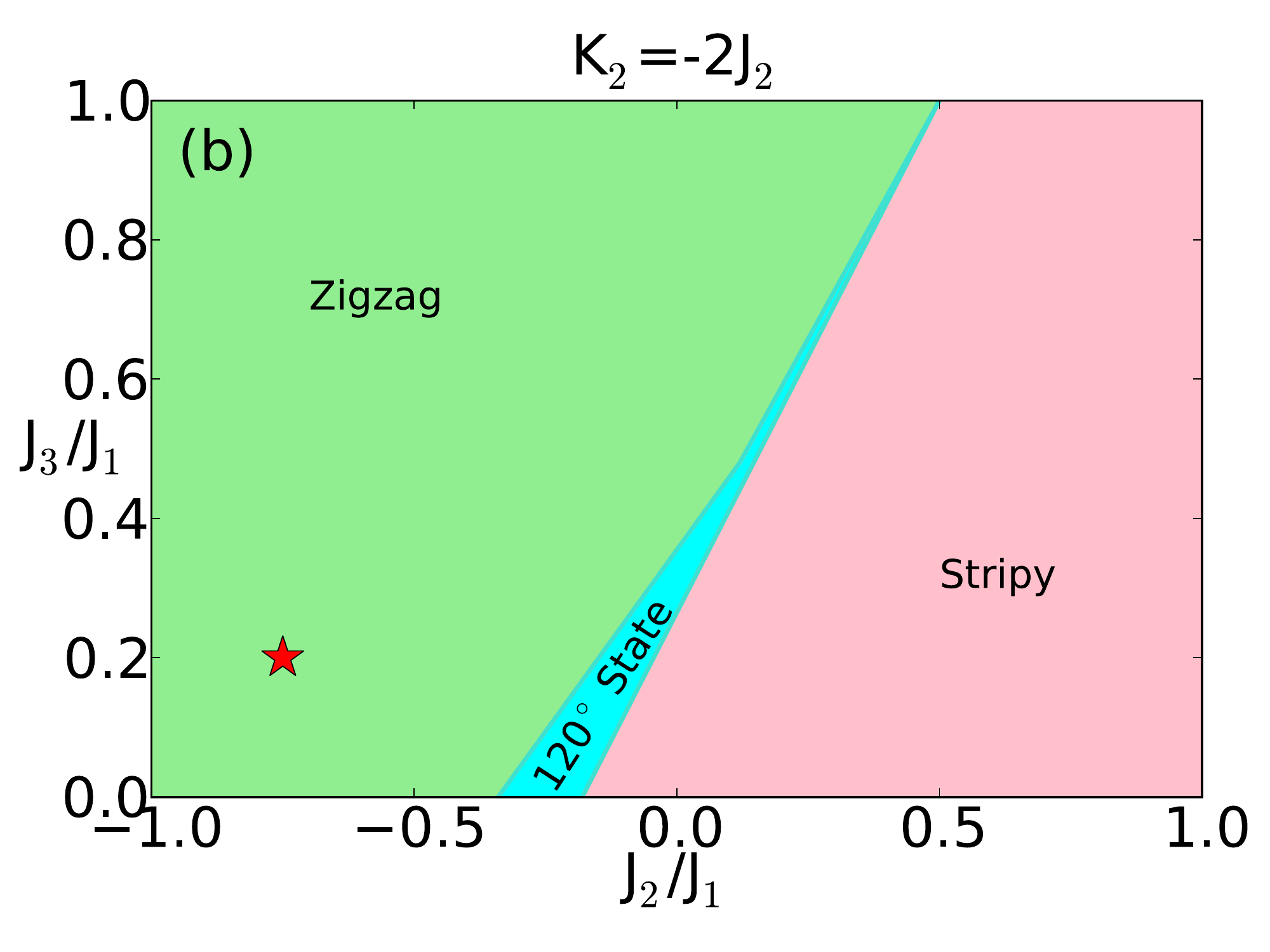}
\includegraphics[width=1.9\columnwidth]{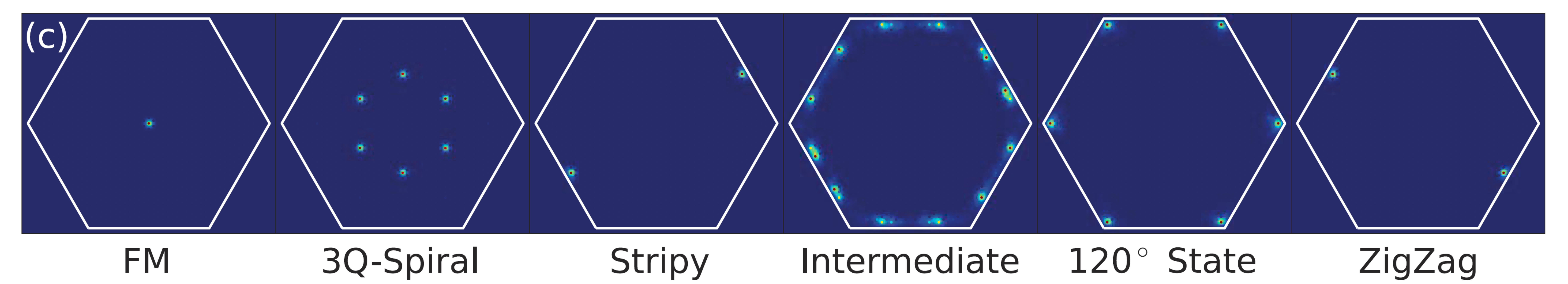}
\caption{(Colors online)
Phase diagrams of the effective model (\ref{minimal}) obtained with the classical Monte Carlo simulations at low temperature $T=0.1 J_1$ for
(a)  second neighbor Kitaev interaction equal to $K_2=0$, (b) second neighbor Kitaev interaction $K_2=-2J_2$.
The simulation is done for $J_1=3$ meV and $K_1=-17$ meV. The blue, rose, green, white, cyan and emerald regions show
the ferromagnetic (FM), the stripy, the zigzag, the incommensurate $3{\mathbf Q}-$spiral, the 120$^\circ$ structure and the intermediate state, respectively. The red star is placed in the region which might well characterize the set of interactions for Na$_2$IrO$_3$. (c) The structure factors
 obtained  as a Fourier transform of a snapshot  of a given configuration for each of these magnetic phases.
 Sharp peaks appear at the corresponding ordering wavevector.
}
\end{figure*}

\subsection{The magnetic phase diagram}

We computed the phase diagram of the  effective model (\ref{minimal}) with  classical Monte Carlo simulations based on the standard Metropolis algorithm.
To explore the physics of the model (\ref{minimal}),  we fix n. n. interactions to $J_1=3$ meV and $K_1=-17$ meV values, which  were obtained by quantum chemistry calculations by Katukuri {\it et al}\cite{katukuri14} and are within the range of parameters obtained by us in this paper. We   compute the phase diagram not only for  ferromagnetic,  $J_2<0$, but also for antiferromagnetic, $J_2>0$, second neighbor interaction. This allows us to compare our findings with other phase diagrams that were previously obtained in the literature.\cite{kimchi11,katukuri14} The simulations were  performed at  low temperature $T=0.1 J_1$, at which for the full range of the considered parameters the model is in the magnetically ordered state.

The  phase diagram   of the model  (\ref{minimal}) in the limit of zero second neighbor Kitaev interaction, $K_2=0$,   is presented in Fig. 5 (a). A more realistic phase diagram computed with $K_2=-2J_2$ is presented in Fig. 5 (b). Even at first glance, we see that the second n. n. Kitaev interaction suppresses the ferromagnetic and spiral phases and stabilizes the antiferromagnetic zigzag and stripy phases.

In order to get a better sense of the basic structure of the different states composing the phase diagrams,  we also performed a numerical Fourier transform of a snapshot of the ground state spin configuration at a given point of the phase diagram.
From that Fourier transform,  we computed the corresponding spin structure factor, which allows us  to determine the dominant wavevectors of that configuration.
We plot  the spin structure factors in Fig. 5 (c).

\subsubsection{ Phase diagram of the $J_1-J_2-J_3-K_1-0$ model (Fig.5 (a)).}

The $K_2=0$ phase diagram is very rich, but overall it  is qualitatively similar to both the classical phase diagram  of the $J_1-J_2-J_3-K_1$\cite{kimchi11,katukuri14} and  of the pure Heisenberg $J_1-J_2-J_3$
model on the honeycomb lattice.\cite{choi12}
 It displays the ferromagnetic (blue region),  the stripy (rose region) and  the zigzag antiferromagnetic states (green region),  the $3{\mathbf Q}-$incommensurate spiral state (white region), the 120$^\circ$ order (cyan region) and a very particular multi-${\mathbf Q}$ incommensurate state (dark cyan region), which we call an "intermediate" phase, as it always separates the 120$^\circ$ order from either the stripy or the zigzag phases. The N\'{e}el antiferromagnetic order is also one of the possible ground states of the model. However,
  the n. n. Kitaev term, $K_1$, and the  second neighbor Heisenberg term, $J_2$, destabilize  it in favor of the stripy and zigzag phases. The N\'{e}el order  is realized only at  values of $J_3/J_1>1$, which are not shown  in the Fig. 5 (a).

The simplest state we find on the phase diagram is  the ferromagnetic state which is characterized by a single ${\mathbf Q}=(0,0)$ wavevector. This state is the ground state in the region  of large ferromagnetic $J_2$ and small $J_3$ couplings. As $J_2$ is decreased  and  $J_3$ is increased, the ferromagnetic state becomes unstable with respect to a spiral state, which is  built out of three  incommensurate wavevectors related by $C_3$ rotation.  Because the ordering ${\mathbf Q}$ vectors are not connected by reciprocal lattice vectors, the spiral phase   represents an example of a $3{\mathbf Q}-$incommensurate order. Note that the magnitude of the ordering wavevector $|{\mathbf Q}|$ varies throughout the phase.

The stripy  and zigzag antiferromagnetic  orders are found for both ferromagnetic and antiferromagnetic $J_2$ interaction of intermediate strength. However, while the stripy order is found at small  values of the third n. n.  interaction, $J_3$,
the  experimentally observed zigzag order  is found only at values  $J_3\ge 0.35 J_1$ which
seem too large given that tight-binding hopping amplitudes
are clearly dominated by the n. n. and the second neighbor terms.\cite{katerina13}
Both the stripy and the zigzag phases  are single-${\mathbf Q}$ orders, characterized by one of the symmetry related wavevectors: ${\mathbf Q}_1=(0,\frac{2\pi}{3})$, ${\mathbf Q}_2=(\frac{\pi}{3},\frac{\pi}{\sqrt{3}})$ and ${\mathbf Q}_3=(-\frac{\pi}{3},\frac{\pi}{\sqrt{3}})$.

 The stripy and the zigzag phases are separated by a 120$^{\circ}$ state characterized by one of the
 ${\mathbf Q}_1=(\frac{4\pi}{3\sqrt{3}},0)$, ${\mathbf Q}_2=(\frac{2\pi}{3\sqrt{3}},\frac{2\pi}{3})$ and ${\mathbf Q}_3=(-\frac{2\pi}{3\sqrt{3}},\frac{2\pi}{3})$
 wavevectors. Because these vectors are connected by the reciprocal lattice vectors, this  is a coplanar  single-${\mathbf Q}$ spiral which describes the 120$^{\circ}$  spin ordering within each  of the two sublattices forming the honeycomb lattice.  As $x$, $y$ and $z$ components of spins are all equally modulated in  this 120$^{\circ}$ state, the spins in this state are lying in  one of the (111) planes.

 The transition from the stripy and the zigzag states into the 120$^{\circ}$ state  is not direct;  it
  happens through the intermediate phase.  This transition can be   understood by looking at the
evolution of the spin structure factors. We find that before the onset of the 120$^{\circ}$ state the transition from a single-${\mathbf Q}$ stripy (or a single-${\mathbf Q}$ zigzag) state to a state  defined by a superposition of three different stripy (zigzag) phases. The structure factor for this state is characterized by the presence of six peaks situated in the middle of the edges of the first BZ hexagon. These peaks  split  into two incommensurate peaks with ${\mathbf Q}$ vectors sliding along the edges (see Fig. 5 (c) for the structure factor corresponding to the Intermediate phase) until they reach wavevectors at the hexagon's corners characterizing the 120$^{\circ}$ structure.
Here, we note that this 120$^{\circ}$ state separating the stripy and the zigzag phases was also obtained  by Rau {\it et al}\cite{rau14} as a classical ground state of the n. n. super-exchange   in the presence of the symmetric off-diagonal exchange.

Here a comment is in order.
In each of the stripy and the zigzag  phases obtained in the Kitaev-Heisenberg models without further neighbor interactions,\cite{jackeli10,jackeli13,price12,price13} the spins were aligned along  one of  the cubic directions. The spin direction was locked to the spatial orientation of a stripy  or a zigzag pattern defined by the wavector ${\mathbf Q}$. Both the  locking of the  spin direction  and the way the  translational symmetry is broken, i.e. the choice of ${\mathbf Q}$, are defined on the classical level.

In the absence of $J_2$ and $J_3$ interactions, the stripy phase is stabilized only for the ferromagnetic n. n. Kitaev interaction, $K_1<0$,  and the zigzag phase is stabilized only for the antiferromagnetic  n. n. Kitaev interaction, $K_1>0$. Consider  the stripy  order with ferromagnetic $z$-bonds.
  In this state,
  the spins and, therefore, the order parameter are pointing along $z$ cubic  axis.  This state has the lowest classical energy, because such a direction of the order parameter maximizes the energy gain   due to the ferromagnetic Kitaev  interaction on ferromagnetic $z$-bonds. The same reasoning explains  why the spins in $x$ and $y$ stripes are pointing along the $x$ and $y$  axes respectively.
  
Next, consider the zigzag  order characterized by ferromagnetic $x-$ and $y-$bonds. 
In this state, the spins  also  point along the $z$ cubic  axis because it maximizes the energy gain   due to the antiferromagnetic Kitaev  interaction on the antiferromagnetic $z-$ bonds.

In the presence of further neighbor couplings  the situation is different. As we can see in Fig. 5 (a),  both the stripy and  the zigzag  order can be stabilized for the ferromagnetic n. n. Kitaev interaction.
While the situation for the stripy phase is the same as before, where  the spins  point along the cubic direction corresponding to the label of the ferromagnetic bond to gain energy from the  ferromagnetic Kitaev interaction,  the direction of the zigzag order parameter is not defined on the classical level.
Instead, there are two ferromagnetic bonds in the zigzag phase, e.g. $x$ and $y$.
 Thus, all zigzag states  characterized by an order parameter pointing along any direction in the $xy$-plane are  classically degenerate. The direction of the order parameter is then selected by order from disorder mechanism,  in which spin fluctuations (quantum or thermal) remove the accidental degeneracy and select the true ordered state. We have checked with Monte Carlo simulations that thermal fluctuations  again choose the states in which spins point along either $x$ or $y$   cubic directions. The full finite-temperature phase diagram for the  model (\ref{minimal}) will be published elsewhere.

\subsubsection{ Phase diagram of the $J_1-J_2-J_3-K_1-K_2$ model (Fig. 5 (b)).}

In Fig. 5 (b), we present the magnetic phase diagram of the model (\ref{minimal}) when the second neighbor Kitaev interaction  is equal to $K_2=-2J_2$, as predicted by our theory when the second neighbors are coupled only through the Ir-O-Na-O-Ir superexchange path. 
We see that the phase diagram greatly simplifies.
 The second neighbor Kitaev term  suppresses  the spiral  and the ferromagnetic phases in favor of the stripy and zigzag order which now dominate for antiferromagnetic and ferromagnetic $J_2$, respectively.
  These two phases are still separated by the 120$^{\circ}$ order and Intermediate phase, but   both the 120$^{\circ}$  phase and, especially, the Intermediate phase  shrink significantly.
However, the most important effect  of the second neighbor Kitaev term is that  for  sufficient ferromagnetic $J_2<0$, it stabilizes the zigzag even for $J_3=0$. In Fig. 5 (b), we put the red star  next to the point which might well characterize the set of interactions for Na$_2$IrO$_3$.

It is worth noting that addition of  non-zero $K_2$ interaction also does not determine the direction of zigzag order parameter on the classical level.  For the zigzag order with antiferromagnetic $z$-bonds discussed above, all states with spins  lying in the $xy$-plane remain classically degenerate. This can be understood as follows.  In the zigzag order  with antiferromagnetic $z$-bonds, the second n. n. ${\tilde z}$-bonds  are ferromagnetic while the ${\tilde x}$- and ${\tilde y}$-bonds  are antiferromagnetic.
 Thus, the antiferromagnetic $K_2$  coupling on these bonds will keep the spins in the $xy$-plane.  However, since there is an equal number of ${\tilde x}$- and ${\tilde y}$-bonds, the $K_2$ interaction  does not lift the classical degeneracy. A particular spin direction, $x$ or $y$, is again chosen by fluctuations.

\section{Conclusions}\label{sec:conclusion}

To summarize, two avenues were explored in this work.
First, we  performed the derivation of an effective super-exchange Hamiltonian
that governs the magnetic properties of the honeycomb iridates treating the  many-body  and single electron interactions on an equal footing.
We demonstrated that  in the presence of strong SO coupling,  this effective Hamiltonian forms a symmetric second-rank tensor with non-equivalent diagonal and non-zero off-diagonal elements. We performed a detailed analysis of the magnetic interactions as a function of  the Hund's coupling representing the electronic correlations  and the trigonal CF splitting which governs the single-electron physics.
We showed that   the main role of the Hund's coupling is that it is responsible for the appearance of the Kitaev anisotropic interactions via the non-equivalence  of the diagonal elements. The
 trigonal CF  also affects the diagonal interactions, however, it's dominating role is in controlling the  strength of  the  off-diagonal interactions. While these interactions might  be significantly increased by external pressure,
at ambient pressure the trigonal CF distortion is small and, consequently, the off-diagonal interactions are subdominant. Thus, we neglected  off-diagonal terms in the  derivation of the super-exchange model  (\ref{minimal}), which we believe is the minimal model to describe the Na$_2$IrO$_3$ compound. This model includes five Ir-Ir couplings:
 n. n. {\it antiferromagnetic}  Heisenberg and {\it ferromagnetic} Kitaev interactions, next n. n. {\it ferromagnetic} Heisenberg and {\it antiferromagnetic} Kitaev interactions, and third n. n. {\it antiferromagnetic} Heisenberg interaction.

The study of the classical phase diagram for this minimal model   constitutes the second part of the paper. We computed  the low temperature phase diagram of the  effective model (\ref{minimal}) with  classical Monte Carlo simulations. Due to the presence of the  anisotropic Kitaev interactions and the frustration introduced by the competition of the spin couplings between n. n. and second neighbors,  the  resulting phase diagram  is very rich. It contains both various commensurate states  and incommensurate  single-${\mathbf Q}$ and multi-${\mathbf Q}$
phases, whose  regions of stability   are controlled by the ratios between competing exchange constants.
 We showed that the second neighbor Kitaev term plays an important role in the stabilization of the  commensurate antiferromagnetic zigzag phase which has been experimentally observed in Na$_2$IrO$_3$. In our simulations, we found this phase to be the ground state for parameters of the model of both the correct signs and magnitudes.

{\it Acknowledgements.}
We thank Katerina Foyevtsova, Michel Gingras, George Jackeli and Arun Paramekanti for useful discussions.
This work was supported, in part, by the National Science Foundation under Grant No.
PHYS-1066293 and the hospitality of the Aspen Center for Physics.
N.P. and Y.S. acknowledge the support from NSF grant DMR-1255544.
  P.W. thanks the Department of Physics at the University of Wisconsin-Madison for hospitality during a stay as a visiting professor. P.W. also acknowledges partial support through the DFG research unit "Quantum phase transitions".

\appendix

\section{The structure of  the  exchange coupling tensor $\Gamma^{\alpha\beta }$}
The  elements of  the  exchange coupling tensor $\Gamma^{\alpha\beta }$ are given by the following expressions:
\begin{eqnarray}
J_{x}=-\sum_{\xi }\frac{1}{\epsilon _{\xi }}&&\Bigl( A_{\uparrow \uparrow
}^{\xi }\left(A_{\downarrow \downarrow }^{\xi }\right)^*+ A_{\downarrow \downarrow
}^{\xi }\left(A_{\uparrow \uparrow }^{\xi }\right)^*\\\nonumber &&+ A_{\uparrow \downarrow
}^{\xi }\left(A_{\downarrow \uparrow }^{\xi }\right)^*+ A_{\downarrow \uparrow
}^{\xi }\left(A_{\uparrow \downarrow }^{\xi }\right)^*\Bigr),
\end{eqnarray}
\begin{eqnarray}
J_{y}=\sum_{\xi }\frac{1}{\epsilon _{\xi }}&&\Bigl( A_{\uparrow \uparrow
}^{\xi }\left(A_{\downarrow \downarrow }^{\xi }\right)^*+ A_{\downarrow \downarrow
}^{\xi }\left(A_{\uparrow \uparrow }^{\xi }\right)^*\\\nonumber&&- A_{\uparrow \downarrow
}^{\xi }\left(A_{\downarrow \uparrow }^{\xi }\right)^*- A_{\downarrow \uparrow
}^{\xi }\left(A_{\uparrow \downarrow }^{\xi }\right)^*\Bigr),
\end{eqnarray}
\begin{eqnarray}
 J_{z}=-\sum_{\xi }\frac{1}{\epsilon _{\xi }}&&\Bigl( A_{\uparrow \uparrow
}^{\xi }\left(A_{\uparrow \uparrow }^{\xi }\right)^*+ A_{\downarrow \downarrow
}^{\xi }\left(A_{\downarrow \downarrow }^{\xi }\right)^*\\\nonumber
&&- A_{\uparrow \downarrow
}^{\xi }\left(A_{\uparrow \downarrow }^{\xi }\right)^*- A_{\downarrow \uparrow
}^{\xi }\left(A_{\downarrow \uparrow }^{\xi }\right)^*\Bigr),
\end{eqnarray}
\begin{eqnarray}
 J_{z}=-\sum_{\xi }\frac{1}{\epsilon _{\xi }}&&\Bigl( A_{\uparrow \uparrow
}^{\xi }\left(A_{\uparrow \uparrow }^{\xi }\right)^*+ A_{\downarrow \downarrow
}^{\xi }\left(A_{\downarrow \downarrow }^{\xi }\right)^*\\\nonumber
&&- A_{\uparrow \downarrow
}^{\xi }\left(A_{\uparrow \downarrow }^{\xi }\right)^*- A_{\downarrow \uparrow
}^{\xi }\left(A_{\downarrow \uparrow }^{\xi }\right)^*\Bigr),
\end{eqnarray}
\begin{eqnarray}
 J_{xy}=\imath\sum_{\xi }\frac{1}{\epsilon _{\xi }}&&\Bigl(
 A_{\uparrow \uparrow
}^{\xi }\left(A_{\downarrow \downarrow }^{\xi }\right)^*- A_{\downarrow \downarrow
}^{\xi }\left(A_{\uparrow \uparrow }^{\xi }\right)^*\\\nonumber
&&+ A_{\downarrow \uparrow
}^{\xi }\left(A_{\uparrow \downarrow }^{\xi }\right)^*- A_{\uparrow \downarrow
}^{\xi }\left(A_{\downarrow \uparrow }^{\xi }\right)^*\Bigr),
\end{eqnarray}
\begin{eqnarray}
 J_{yx}=\imath\sum_{\xi }\frac{1}{\epsilon _{\xi }}&&\Bigl(
 A_{\uparrow \uparrow
}^{\xi }\left(A_{\downarrow \downarrow }^{\xi }\right)^*- A_{\downarrow \downarrow
}^{\xi }\left(A_{\uparrow \uparrow }^{\xi }\right)^*\\\nonumber
&&- A_{\downarrow \uparrow
}^{\xi }\left(A_{\uparrow \downarrow }^{\xi }\right)^*+ A_{\uparrow \downarrow
}^{\xi }\left(A_{\downarrow \uparrow }^{\xi }\right)^*\Bigr),
\end{eqnarray}
\begin{eqnarray}
 J_{xz}=\sum_{\xi }\frac{1}{\epsilon _{\xi }}&&\Bigl(
 A_{\uparrow \uparrow
}^{\xi }\left(A_{\downarrow \uparrow }^{\xi }\right)^*- A_{\downarrow \downarrow
}^{\xi }\left(A_{\uparrow \downarrow }^{\xi }\right)^*\\\nonumber
&&+ A_{\downarrow \uparrow
}^{\xi }\left(A_{\uparrow \uparrow }^{\xi }\right)^*- A_{\uparrow \downarrow
}^{\xi }\left(A_{\downarrow \downarrow }^{\xi }\right)^*\Bigr),
\end{eqnarray}
\begin{eqnarray}
 J_{zx}=\sum_{\xi }\frac{1}{\epsilon _{\xi }}&&\Bigl(
 A_{\uparrow \uparrow
}^{\xi }\left(A_{\uparrow \downarrow }^{\xi }\right)^*- A_{\downarrow \downarrow
}^{\xi }\left(A_{\downarrow \uparrow }^{\xi }\right)^*\\\nonumber
&&+ A_{\uparrow \downarrow
}^{\xi }\left(A_{\uparrow \uparrow }^{\xi }\right)^*- A_{\downarrow \uparrow
}^{\xi }\left(A_{\downarrow \downarrow }^{\xi }\right)^*\Bigr),
\end{eqnarray}
\begin{eqnarray}
 J_{yz}=\imath\sum_{\xi }\frac{1}{\epsilon _{\xi }}&&\Bigl(
 A_{\uparrow \uparrow
}^{\xi }\left(A_{\downarrow \uparrow }^{\xi }\right)^*+ A_{\downarrow \downarrow
}^{\xi }\left(A_{\uparrow \downarrow }^{\xi }\right)^*\\\nonumber
&&- A_{\uparrow \downarrow
}^{\xi }\left(A_{\downarrow \downarrow }^{\xi }\right)^*- A_{\downarrow \uparrow
}^{\xi }\left(A_{\uparrow \uparrow }^{\xi }\right)^*\Bigr),
\end{eqnarray}
\begin{eqnarray}
 J_{zy}=\imath\sum_{\xi }\frac{1}{\epsilon _{\xi }}&&\Bigl(
 A_{\uparrow \uparrow
}^{\xi }\left(A_{\uparrow \downarrow }^{\xi }\right)^*+ A_{\downarrow \downarrow
}^{\xi }\left(A_{\downarrow \uparrow }^{\xi }\right)^*\\\nonumber
&&- A_{\uparrow \downarrow
}^{\xi }\left(A_{\uparrow \uparrow }^{\xi }\right)^*- A_{\downarrow \uparrow
}^{\xi }\left(A_{\downarrow \downarrow }^{\xi }\right)^*\Bigr),
\end{eqnarray}
Here, in order to shorten notations, we omitted the site indices denoting
$A_{n,n^{\prime};\sigma ,\sigma ^{\prime }}^{\xi }\equiv A_{\sigma ,\sigma ^{\prime }}^{\xi }$ and $A_{n^{\prime},n;\sigma^{\prime} ,\sigma }^{\xi }\equiv \left(A_{\sigma ,\sigma ^{\prime }}^{\xi }\right)^*$.


\begin{thebibliography}{100}
\bibitem{comment}
 The paths which connect third nearest neighbors are similar to the second neighbors. Starting from the same Ir$^{4+}$ ions as in the main text, the four dominant paths connecting third neighbors will be $
{\rm Ir}\,(Y)\rightarrow
{\rm O}\,(p_z)\rightarrow
{\rm Na}\,(s)\rightarrow
{\rm O}\,(p_z)\rightarrow
{\rm Ir}\,(Y)$, $
{\rm Ir}\,(Y)\rightarrow
{\rm O}\,(p_x)\rightarrow
{\rm Na}\,(s)\rightarrow
{\rm O}\,(p_x)\rightarrow
{\rm Ir}\,(Y)$, $
{\rm Ir}\,(Y)\rightarrow
{\rm O}\,(p_x)\rightarrow
{\rm Na}\,(s)\rightarrow
{\rm O}\,(p_z)\rightarrow
{\rm Ir}\,(Y)$, $
{\rm Ir}\,(Y)\rightarrow
{\rm O}\,(p_z)\rightarrow
{\rm Na}\,(s)\rightarrow
{\rm O}\,(p_x)\rightarrow
{\rm Ir}\,(Y)$. All these paths connect the same $Y$ orbitals and, thus, contribute mostly to the isotropic Heisenberg exchange. The small anisotropic part is similar in form to the $J_1^{xy}$ term.
\end{thebibliography}
\end{document}